\documentstyle[11pt]{article}
\begin{document}

\title{Quantum Symmetry\\{\normalsize
An Approach to Quantum Gauge Symmetry via Reconstruction Theorems}}
\author{Reinhard H\"aring\\Institut f\"ur angewandte Mathematik\\
        c/o Lersnerstr. 19\\60322 Frankfurt\\Germany}
\date{October, 5, 1993}
\maketitle

\begin{abstract}
The representations of the observable algebra of a
low dimensional quantum field theory form the objects
of a braided tensor category. The search for gauge symmetry
in the theory amounts to finding an algebra which has the
same representation category.

In this paper we try to establish that every quantum field theory
satisfying some basic axioms
posseses a weak quasi Hopf algebra as gauge symmetry.
The first step is to construct a functor from the representation category
to the category of finite dimensional vector spaces.
Given such a functor we can use a generalized reconstruction theorem
to find the symmetry algebra.
It is shown how this symmetry algebra is used to build a
gauge covariant field algebra and we investigate the question why this
generality is necessary.
\end{abstract}

%

\newcommand{\EE}{\mathop{\rm I\! E}\nolimits}
\newcommand{\E}{\mathop{\rm E}\nolimits}
\newcommand{\I}{\mathop{\rm Im\, }\nolimits}
\newcommand{\Str}{\mathop{\rm Str\, }\nolimits}
\newcommand{\Sdet}{\mathop{\rm Sdet\, }\nolimits}
\newcommand{\STr}{\mathop{\rm STr\, }\nolimits}
\newcommand{\R}{\mathop{\rm Re\, }\nolimits}
\newcommand{\CC}{\mathop{\rm C\!\!\! I}\nolimits}
\newcommand{\FF}{\mathop{\rm I\! F}\nolimits}
\newcommand{\KK}{\mathop{\rm I\! K}\nolimits}
\newcommand{\LL}{\mathop{\rm I\! L}\nolimits}
\newcommand{\MM}{\mathop{\rm I\! M}\nolimits}
\newcommand{\NN}{\mathop{\rm I\! N}\nolimits}
\newcommand{\PP}{\mathop{\rm I\! P}\nolimits}
\newcommand{\QQ}{\mathop{\rm I\! Q}\nolimits}
\newcommand{\RR}{\mathop{\rm I\! R}\nolimits}
\newcommand{\ZZ}{\mathop{\sl Z\!\!Z}\nolimits}
\newcommand{\integer}{\mathop{\rm int}\nolimits}
\newcommand{\erf}{\mathop{\rm erf}\nolimits}
\newcommand{\diag}{\mathop{\rm diag}\nolimits}
\newcommand{\fl}{\mathop{\rm fl}\nolimits}
\newcommand{\eps}{\mathop{\rm eps}\nolimits}
\newcommand{\var}{\mathop{\rm var}\nolimits}


\newcommand{\pfeil}{\rightarrow}

\newcommand{\kat}{{\cal C}}
\newcommand{\rmat}{{\cal R}}
\newcommand{\oalg}{{\cal A}}
\newcommand{\falg}{{\cal F}}
\newcommand{\eich}{{\cal G}}
\newcommand{\hilb}{{\cal H}}
\newcommand{\calm}{{\cal M}}
\newcommand{\mod}{{\cal M}}
\newcommand{\kegel}{{\cal O}}
\newcommand{\kegels}{{\cal K}}
\newcommand{\bigrho}{\rho_\oplus}
\newcommand{\bigphi}{\phi_\oplus}

\newcommand{\tprod}{\stackrel{\sim}{\otimes}}
\newcommand{\iso}{\stackrel{\sim}{=}}
\newcommand{\usw}{u.s.w.}
\newcommand{\bzw}{b.z.w.}

\newcommand{\quer}[1]{\overline{#1}}
\newcommand{\schlange}[1]{\widetilde{#1}}
\newcommand{\CVO}[3]{{{#3 \choose #1\hspace{3pt}#2}}}
\newcommand{\clebsch}[6]{{\left[{#1\atop#4}\hspace{3pt}{#2\atop#5}
                                     \hspace{3pt}{#3\atop#6}\right]}}
\newcommand{\sjsymbol}[6]{{\left\{{#1\atop#4}\hspace{3pt}{#2\atop#5}
                                     \hspace{3pt}{#3\atop#6}\right\}}}
\newcommand{\ket}[1]{|#1\rangle}
\newcommand{\vak}{\ket{0}}
\newcommand{\bild}[3]{{        
  \unitlength1mm
  \begin{figure}[ht]
  \begin{picture}(120,#1)\end{picture}
  \caption{\label{#3}#2}
  \end{figure}
}}

\newenvironment{bew}{Proof:\small}{\hfill$\Box$}

\newtheorem{bem}{Note}
\newtheorem{bsp}{Example}
\newtheorem{axiom}{Axiom}
\newtheorem{de}{Definition}
\newtheorem{satz}{Proposition}
\newtheorem{lemma}[satz]{Lemma}
\newtheorem{kor}[satz]{Corollary}
\newtheorem{theo}[satz]{Theorem}

\newcommand{\sbegin}[1]{\small\begin{#1}}
\newcommand{\send}[1]{\end{#1}\normalsize}

\sloppy

{\small \tableofcontents}

\section{Introduction}

The structure of quantum field theories depends sensitive on the dimension
of space time. It affects statistics and coupled with it gauge symmetry.
It is by now well established (at least in the case of charges that
can be localized in spacelike cones) that
in four and more dimensions only permutation group
statistics (i.e. Bose and Fermi statistics) is possible.
In two space time dimension braid group statistic rules the exchange of
operators. Between this antipodes fall three dimensional models which may
have permutation or braid group statistics depending on the
localization of charges.
'Dual' to statistics is the notion of gauge symmetry.
It was shown by Doplicher
and Roberts \cite{DR1} that all field theories with
permutation group statistics
posess a (uniquely determined) compact gauge group. Until now a comparable
result for braid group statistics has been lacking. Quantum groups
(quasitriangular
Hopf algebras) were supposed to replace the compact gauge groups.
However it was soon realized that they have more representations
than needed to serve as a gauge algebra. In one way or another one had
to abandon these unphysical (indecomposable) representations and keep
only the physical (fully decomposable) ones. The majority of
researchers decided
to accomplish this truncation simply by forgeting  about them.
While studying the Ising model Mack and Schomerus
\cite{ms0} noticed that this leads to contradictions. They introduced
weak quasi Hopf algebras in \cite{ms1} which are not plagued by
unphysical representations
(because truncation is build into their coproduct)
and showed how they can be used to build
a gauge covariant field algebra for the Ising model \cite{ms2}.

But still the situation was unsatisfactory. The Ising model remained
the only example where the gauge algebra was explicitly known.
A systematic procedure for constructing the gauge algebra was needed.
Already Doplicher and Roberts had used categorial techniques.
They realized that the relevant condition a gauge algebra must
fulfil is that its representation category (the representations
are the objects in this category and the intertwiners between them are the
morphisms) is equivalent to the representation category of the
observable algebra. Doplicher and Roberts were able to solve this problem,
the reconstruction of an algebraic object from its representation theory,
in the classical case. For braided categories Majid proved a reconstruction
theorem. Given a (quasi) tensor functor from the category
to the vector spaces
this theorem reconstructs a (quasi) Hopf algebra. In the physical
relevant cases
such functors can't exist because of the need for truncation.
Kerler \cite{kerler} had the idea to generalize the reconstruction theorem
to the case of weak quasi tensor functors (to allow truncation). He
supposed that the reconstructed algebra would be a
weak quasi Hopf algebra in the sense
of Mack and Schomerus. However he was not able to give a proof and he guessed
wrong requirements for the functor. Furthermore he
couldn't give a construction
of the needed weak quasi tensor functor.
In this paper we repair this situation.
We show that weak quasi tensor functors always exist
and proof a generalized
version of Majid's reconstruction theorem that allows the construction
of weak quasi Hopf algebras. We carry out the
reconstruction explicitly and find
almost (The antipode diverges.)  the same algebra as
Mack and Schomerus constructed by hand.
In contrast to the classical situation analysed
by Doplicher/Roberts we find that
the gauge algebra is not uniquely determined.

The next logical step is the construction of gauge covariant field algebra.
This problem has already been attacked by many authors
\cite{moorecomm}, \cite{ms0}, \cite{rehren3}, \cite{gomez2}, \cite{ffk2}.
Our construction is generalizing\footnote{After the construction of the
field algebra was completed we received a copy of Schomerus' PhD thesis
\cite{schomerus1}. He now has a field algebra that is as general as ours. We
hope that our alternative construction will still be useful.}
and correcting these works into a
powerful construction that is model independent and close to physical
intuition. The fields form an involutive algebra which acts on a
Hilbert space with positive definite scalar product. The
construction is symmetric between the gauge and the observable
algebra.

We prove braid und fusion relations in this field algebra.
The field operators
obey braid relations with the $\rmat$ matrix of the gauge algebra. This
shows the deep interplay between symmetry and statistics.

Some material about ultra weak quasi Hopf algebras is spread across the paper
which we suppose to be the right framework for the construction of
chiral algebras (observable algebras).

\section{Braided Tensor Categories}

The occurence of representations with braiding properties
in low dimensional quantum field theories
motivates the introduction of generalized tensor categories
which are not symmetric.

\subsection{Definitions}

The objects of a category $\kat$ are denoted by $X\in{\rm Obj}(\kat)$,
the morphisms  between
$X,Y\in {\rm Obj}(\kat)$ with ${\rm Mor}(X,Y)$.
${\rm End}(X):={\rm Mor}(X,X)$.

\begin{de}[Monoidal category, braided tensor category]
A category $\kat$ is called {\em\bf monoidal}  if
there is a functor $\tprod:\kat\times\kat\pfeil\kat$ obeying:
\begin{enumerate}
\item There are functorial isomorphisms:
\begin{equation}
\Phi_{X,Y,Z}:X\tprod(Y\tprod Z)\stackrel{\sim}{\longrightarrow}
(X\tprod Y)\tprod Z
\qquad X,Y,Z\in {\rm Obj}(\kat)
\end{equation}
$\Phi$ satisfies the following  {\em\bf pentagon identity}:
\begin{equation}
\begin{array}{lcccr}
X\tprod(Y\tprod(Z\tprod T))&\stackrel{\Phi}{\longrightarrow}&
(X\tprod Y)\tprod(Z\tprod T)&\stackrel{\Phi}{\longrightarrow}&
((X\tprod Y)\tprod Z)\tprod T\\
\downarrow1\tprod\Phi&&&&\uparrow\Phi\tprod1\\
X\tprod((Y\tprod Z)\tprod T)&&\stackrel{\Phi}{\longrightarrow}&&
(X\tprod(Y\tprod Z))\tprod T
\end{array}\end{equation}
$\Phi$ is called the {\em\bf assoziator}.
\item There is an {\em\bf identity object} $1\in {\rm Obj}(\kat)$, such that
$r_X:X\mapsto 1\tprod X$ and $l_X:X\mapsto X\tprod1$ are
equivalences of categories compatible with $\Phi$:
\begin{equation}
\Phi_{1,X,Y}\circ l_{X\tprod Y}=l_X\tprod id_Y
\end{equation}
$r$ satisfies analog identities.
\end{enumerate}
A monidal category is called {\em\bf braided tensor category}
if there is a functorial isomorphism $\Psi$ such that:
\begin{equation}
\Psi_{X,Y}:X\tprod Y\stackrel{\sim}{\longrightarrow}Y\tprod X
\qquad X,Y\in {\rm Obj}(\kat)
\end{equation}
$\Psi$ satisfies two
{\em\bf hexagon identities} and
is compatible with $l$ and $r$.
\begin{equation}\begin{array}{lcccr}
X\tprod(Y\tprod Z)&\stackrel{\Phi}{\longrightarrow}&(X\tprod Y)\tprod Z&
\stackrel{\Psi}{\longrightarrow}&Z\tprod(X\tprod Y)\\
\downarrow1\tprod\Psi&&&&\downarrow\Phi\\
X\tprod(Z\tprod Y)&\stackrel{\Phi}{\longrightarrow}&(X\tprod Z)\tprod Y&
\stackrel{\Psi\tprod1}{\longrightarrow}&(Z\tprod X)\tprod Y
\end{array}\end{equation}
\begin{equation}\begin{array}{lcccr}
(X\tprod Y)\tprod Z&\stackrel{\Phi^{-1}}{\longrightarrow}&X\tprod(Y\tprod Z)&
\stackrel{\Psi}{\longrightarrow}&(Y\tprod Z)\tprod X\\
\downarrow\Psi\tprod1&&&&\downarrow\Phi^{-1}\\
(Y\tprod X)\tprod Z&\stackrel{\Phi^{-1}}{\longrightarrow}&Y\tprod(X\tprod Z)&
\stackrel{1\tprod\Psi}{\longrightarrow}&Y\tprod(Z\tprod X)
\end{array}\end{equation}
\begin{equation}\begin{array}{lcr}
X\tprod Y&\stackrel{1\tprod r}{\longrightarrow}&X\tprod(Y\tprod1)\\
\downarrow r\tprod1&&\downarrow1\tprod\Psi\\
(X\tprod1)\tprod Y&\stackrel{\Phi}{\longrightarrow}&X\tprod(1\tprod Y)
\end{array}\end{equation}

Furthermore we demand $l_X=\Psi\circ r_X$.
If $\Psi_{X,Y}\Psi_{Y,X}=id$, $\kat$ is called {\em\bf tensor category}
or (in contrast to braided  tensor categories)
{\em\bf symmetric tensor category}.
\end{de}

We assume all categories to be abelian (and all functors to additive)
with direct sum $\schlange{\oplus}$ and
zero element $0$.

\begin{de}
\begin{enumerate}
\item $X\in\kat$ is called
{\em\bf indecomposable} if
Mor$(X,X)=\mbox{span id}_X\oplus {\cal N}$ where ${\cal N}$ consists only
of nilpotent elements.
$X$ is called {\em\bf irreducible},
if ${\cal N}=0$. The set of irreducible objects is denoted by
{\bf ${\rm Obj}_{irr}$}.
$\kat$ is called {\em\bf fully reducible},
if all $X\in{\rm Obj}(\kat)$ are isomorph to sums of irreducible objects.
A fully reducible category is called {\em\bf Schur category}\footnote{We thank
A. Schmidt for pointing out the importance of this property.} if
there are no morphisms between inequivalent irreducible
objects.\footnote{Categories of
representation are by Schur's lemma Schur categories.}

In a {\em\bf rational category} there exist only finitely many equivalence
classes
of indecomposable objects. In a {\em\bf quasi-rational category} every object
is isomrphic to a finite sum of indecomposable objects.

$\kat$ is called {\em\bf irredundant},
if $X\iso Y\Rightarrow X=Y$.

A local finite (i.e. $\forall X,Y\in{\rm Obj}(\kat)
{\rm dim}({\rm Mor}(X,Y))<\infty$),
abelian  braided tensor category is called {\em\bf semisimple},
if Mor$(X,X)$ is
a semisimple algebra for all objects.
\item In a {\em\bf $C^\ast$-category} $\kat$
all ${\rm Mor}(X,Y)$ are normed $\CC$-vector spaces with an
antilinear involution $\dagger:{\rm Mor}(X,Y)\pfeil {\rm Mor}(Y,X)$
such that $(fg)^\dagger=g^\dagger f^\dagger,
||f^\dagger f||=||f||^2$. ${\rm Mor}(X,X)$
is then a unital $C^\ast$-algebra.
Fully reducible $C^\ast$ categories are Schur categories (Assume
$f\in{\rm Mor}(X,Y)$ to be a morphism between inequivalent, irreducible
objects. $f^\dagger f\in{\rm Mor}(X,X)$  and $ff^\dagger\in{\rm Mor}(Y,Y)$.
Just as $f$ these maps can't be isos so they vanish.
 $0=||f^\dagger f||=||f||^2\Rightarrow f=0$).
\item A functor $F:\kat_1\pfeil \kat_2$ is called {\rm\bf faithful} if
$F:Mor(X,Y)\pfeil Mor(F(X),F(Y)) $
is injective for all $X,Y\in{\rm Obj}(\kat_1)$.
\end{enumerate}\end{de}

By Wedderburn's theorem we get for locally finite categories:
\begin{lemma}\label{halbred}
$X\in{\rm Obj}(\kat)$ fully reducible  $\Longleftrightarrow$
${\rm End}(X)$ semisimple.

$\kat$ fully reducible $\Longleftrightarrow$
$\kat$ semisimple
\end{lemma}

\begin{de}[Monoidal functor]
A functor $F:\kat_1\pfeil\kat_2$ between two monoidal
categories is called
{\em\bf monoidal} (resp. {\em\bf weakly monoidal})
if there is a functorial isomorphism (resp. epimorphism) $c_{X,Y}$
\begin{equation}
c_{X,Y}:F(X)\tprod_2F(Y)\stackrel{\sim}{\longrightarrow}F(X\tprod_1Y)
\end{equation}
such that $F$ becomes compatible with the associator and the unit:
\begin{equation}\label{assod}\begin{array}{lcccr}
F(X)\tprod(F(Y)\tprod F(Z))&\stackrel{1\tprod c}{\longrightarrow}&
F(X)\tprod F(Y\tprod Z)&\stackrel{c}{\longrightarrow}&F(X\tprod(Y\tprod Z))\\
\downarrow\Phi_2&&&&\downarrow F(\Phi_1)\\
(F(X)\tprod F(Y))\tprod F(Z)&\stackrel{c\tprod1}{\longrightarrow}&
F(X\tprod Y)\tprod F(Z)&\stackrel{c}{\longrightarrow}&F((X\tprod Y)\tprod Z)
\end{array}\end{equation}
\begin{equation} l_2|_{F({\rm Obj}(\kat_1))}=
  c^{-1}\circ F(l_1):F(X)\mapsto F(1)\tprod_2 F(X)
  \iso1\tprod_2 F(X)                        \label{einskomp}
\end{equation}
A functor between two (braided) tensor categories is called
{\em\bf symmetric}  if it is
compatible with the braid isomorphism, i.e.
for all $X,Y\in{\rm Obj}(\kat)$ the diagram
\begin{eqnarray}
F(X)\tprod F(Y)&\stackrel{c}{\longrightarrow}&F( X\tprod Y)\nonumber\\
\downarrow {\scriptstyle \Psi_2} &   &
\downarrow {\scriptstyle F(\Psi_1)}\nonumber\\
F(Y)\tprod F(X)&\stackrel{c}{\longrightarrow}&
F(Y\tprod X)
\label{zopfkomp}
\end{eqnarray}
is commutative.
A monoidal functor between   braided tensor categories is called
a {\em\bf tensor functor}\footnote{This property follows in all cases
with exception
of the ultraweak case from the other axioms. One could therefore
formulate most of the present
paper using only the term monoidal functor.} if:
\begin{equation}
F(\Psi(X\tprod Y))\iso F(X\tprod Y) \label{zopfkomp2}
\end{equation}
In case  (\ref{assod})  is not required $F$ is called
{\em\bf quasi tensor functor} and if
$c_{X,Y}$ is only an epimorphism (but with $c_{X,1}$ and $c_{1,X}$
remaining isomorphisms)
with right inverse $c_{X,Y}^{-1}$ then $F$ is only a
{\em\bf weak quasi tensor functor}.
Finally $F$ is called {\em\bf ultra weak quasi tensor functor}
if (\ref{einskomp})
and $c_{1,X},c_{X,1}$ iso is not postulated but
$c_{1,X}=c_{X,1}\circ\Psi^{\kat_2}$.

If $\kat_1$ and $\kat_2$ are rigid (see below)
then we demand\footnote{In the
non-weak cases the existence of such isomorphisms
follwos from rigidity by
setting $F(X)^\ast:=F(X^\ast), ev_{F(X)}:=F(ev_X)\circ c$.
\cite{majid11}} in addition
the existence of functorial isomorphisms $d_X:F(X)^\ast\pfeil F(X^\ast)$.
\end{de}

$F:\kat\pfeil\underline{Set}$ is called representable, if
$\exists X\in {\rm Obj}(\kat)$ such that $F$ is naturally isomorph to
$Y\mapsto {\rm Mor}(X,Y)$. $X$ is called representing object.

\begin{de}[Internal Hom] One says that the
(braided) tensor
category $\kat$ has an internal Hom, if $\forall X,Y\in {\rm Obj}(\kat)$
the functor $Z\mapsto {\rm Mor}(Z\tprod X,Y)$ is representable.
The representing object
is called  ${\rm Hom}(X,Y)\in {\rm Obj}(\kat)$.
So we have the following functorial isomorphism:
\begin{equation} \label{inthom} {\rm Mor}(Z\tprod X,Y)\iso
{\rm Mor}(Z,{\rm Hom}(X,Y))\end{equation}
$X^\ast:={\rm Hom}(X,1)$ is called the dual object.
\end{de}

Setting $Z=1$ yields ${\rm Mor}(X,Y)\iso {\rm Mor}(1,{\rm Hom}(X,Y))$.

In the  special case  $Z={\rm Hom}(X,Y)$  the
morphism which is mapped to the identity in
${\rm Mor}({\rm Hom}(X,Y),{\rm Hom}(X,Y))$ is denoted by
$ev_{X,Y}:{\rm Hom}(X,Y)\tprod X\pfeil Y$
and called the {\em\bf evaluation}.
$ev_X$ is a shorthand for $ev_{X,1}$.
Dually we have the notion of coevaluation
$coev_X\in{\rm Mor}(1,X\tprod X^\ast)$
characterized by the dual relations
\begin{equation} (ev_X\tprod id)(id\tprod coev_X)=id_{X^\ast}\qquad
 (id\tprod ev_X)(coev_X\tprod id)=id_X \label{coeveq}
 \end{equation}

In rigid categories the existence of $coev$ ist
guaranteed.
\begin{de}[Rigid tensor category] A (braided) tensor category $\kat$
with  internal
Hom is called {\em\bf rigid},
if there are the following functorial isomorphisms
\footnote{Equations (\ref{eq005}) and (\ref{inthom}) allow to dertemine
$(A\tprod B)^\ast=A^\ast\tprod B^\ast$ by setting $Y=1,X=A\tprod B$.}:
\begin{eqnarray}
{\rm Hom}(X_1,Y_1)\tprod {\rm Hom}(X_2,Y_2)\stackrel{\sim}{\longrightarrow}
{\rm Hom}(X_1\tprod X_2,Y_1\tprod Y_2)\label{eq005}\\
X\stackrel{\sim}{\longrightarrow}X^{\ast\ast}
\end{eqnarray}
\end{de}

Setting $Y_1=X_2=1, X_1=X, Y_2=Y$ and using  (\ref{eq005}) one obtains
${\rm Hom}(X,Y)\iso {\rm Hom}(X,1)\tprod {\rm Hom}(1,Y)\iso X^\ast\tprod Y$
(The last isomorphism follows from (\ref{inthom}) for $X=1$.).

\subsection{Semisimple and Irredundant Quasi Tensor Categories}	\label{semi}

In the application we have in mind the categories are
representation categories of algebras.

For an algebra $A$ we let ${\rm Rep}(A)$ denote
its {\em\bf representation category}.
The objects are the  representations
(eventually only special representations are considered,
see below))
of $A$ and the morphisms are the intertwiners.

${\rm Rep}(A)$ is a braided tensor category  if $A$ permits products
of representations which are symmetric upto isomorphisms.

As in representation theory we want to abandon the complications
of equivalent but
yet distinct objects.

\begin{bem}
On can associate to $\kat$ a irredundant category $[\kat]$ which
is by definition
a full subcategory of $\kat$ containing one object in every equivalence
class.
$[\kat]$ is called a skeleton of $\kat$ in \cite{macl}.
\end{bem}

\begin{bem}
Kerler \cite{kerler} showed that to every rigid braided tensor category
$\kat$
there exists  a
canonically associated semisimple quotient braided tensor category
${\cal S}(\kat)$. This construction is compatible with the elimination
of redundancy \cite{haring}:
 $[{\cal S}(\kat)]={\cal S}([\kat])$
\end{bem}

\subsubsection{Description of semisimple categories via
polynomial equations}\label{peq}

Let $\kat$ denote a semisimple braided
Schur tensor category and let $\nabla_0\subset{\rm Obj}_{irr}(\kat)$
contain one object per
irreducible equivalence class.
For each triple $X,Y,Z\in\nabla_0$ let
$N_{X,Y}^Z$ denote the dimension
of ${\rm Mor}(X\tprod Y,Z)$ and choose a basis
$\phi(e)\in{\rm Mor}(X\tprod Y,Z)$
($e=^i\CVO{X}{Y}{Z}$ is a multi index with
$i\in\{1,\ldots,N_{X,Y}^Z\}$.).
$\phi^i\CVO{X}{Y}{Z}\circ\Psi_{Y,X}\in{\rm Mor}(Y\tprod X,Z)$ and
$\phi^i\CVO{X}{M}{R}\circ(id_X\tprod\phi^j\CVO{Y}{Z}{M})
\in{\rm Mor}(X\tprod(Y\tprod Z))$
can then be expanded in the base via matrices
\begin{eqnarray}
\phi(e)\circ\Psi=\sum_f\Omega_{e,f}\phi(f)\\
\phi(e_2)(id\otimes\phi(e_1))(h_2\otimes h_1\otimes h_s)&=&
\sum_{e,f}F_{e_1,e_2;f,e}\phi(e)(\phi(f)\otimes id)
(h_2\otimes h_1\otimes h_s)
\end{eqnarray}
It follows straightforward from the axioms of braided tensor categories that
these matrices satisfy the Moore/Seiberg polynomial equations.

Moore/Seiberg have shown (\cite{moores2},\cite{kratz}) that in
the opposite direction every solution to their equations yields
such a category. Their construction is essentially the following:
Take a set of irreducible objects $X_i,i\in{\cal I}$ and set
${\rm Mor}(X_i,X_j):=\CC\delta_{i,j}id_{X_i}$. Tensor products
are formally introduced via
$X_i\tprod X_j:=\bigoplus_l V_{i,j}^l\otimes X_l$ where $V_{i,j}^l$ are
$N_{i,j}^l$ dimensional vector spaces of morphisms ${\rm Mor}
(X_i\tprod X_j,X_l)$.
The braid iso operates
on this tensor product via the operation of $\Omega$ on $V_{i,j}^l$.

\subsubsection{Construction of (weak) Quasi Tensor Functors}
\label{ktab007}

Let $\nabla_0\subset{\rm Obj}_{irr}(\kat)$ contain one object per
irreducible equivalence class. $\nabla_0^\ast$ can be different
from $\nabla_0$	but it shares the same properties since
the dual of an irreducible (indecomposable)
object is again irreducible (indecomposable).

\begin{de} A function defined on the irreducible objects of a
semisimple, rigid braided Schur tensor category
$D:{\rm Obj}_{irr}(\kat)\pfeil\NN$ which is
constant on equivalence classes is called
{\em\bf weak dimension function}, if:
\begin{equation}
D(1)=1, D(X)=D(X^\ast), D(X)D(Y)\geq\sum_{Z\in\nabla_0}
D(Z){\rm dim}({\rm Mor}(X\tprod Y,Z))
\end{equation}
$D$ is called {\em\bf dimension function} if equality holds.
\end{de}

Dimension functions allow according to ideas of Kerler
\footnote{In the weak case Kerler's idea is wrong:
His $C_{i,j}$ can't be choosen as epimorphisms as required.
The Ising model is a counterexample.}
\cite{kerler}
the construction of  functors:

\begin{satz}\label{funktorkon}
Let $\kat$ be a quasi-rational semisimple,
rigid braided Schur tensor category and $D:{\rm Obj}(\kat)\pfeil\NN$
a (weak) dimension function.
Then there is a faithful (weak) quasi tensor functor
$F:\kat\pfeil\underline{{\rm Vec}}$ into the category of
finite dimensional vectorspaces.
\end{satz}
\begin{bew}
For $X\in\nabla_0$ let $F(X):=\CC^{D(X)}$
and for arbitrary objects $Y\in{\rm Obj}(\kat)$ this is extended via
$F(Y):=\bigoplus_{X\in\nabla_0}{\rm Mor}(X,Y)\otimes F(X)$.
$F$ acts on  morphisms $f\in{\rm Mor}(Y_1,Y_2)$ as
$F(f)\in{\rm Mor}(F(Y_1),F(Y_2))$.
Because of linearity, $F(f)$ needs only be defined on the
summands of type
${\rm Mor}(X,Y_1)\otimes F(X)$.
Let $F(f)(g\otimes x):=f\circ g\otimes x,x\in F(X),g\in{\rm Mor}(X,Y_1)$

Assume $f_1,f_2\in{\rm Mor}(Y_1,Y_2), F(f_1)=F(f_2)$.
By the definition of $F$ this implies
$\forall X\in\nabla_0\forall g\in{\rm Mor}(X,Y_1) f_1\circ g=f_2\circ g$.
Since $\kat$ is assumed to be semisimple we have an isomorphism
$\phi\in{\rm Mor}(Y_1,X_{i_1}\oplus\ldots\oplus X_{i_n}), X_{i_l}\in\nabla_0$.
 From this we get $p_{i_l}\in{\rm Mor}(Y_1,X_{i_l}), q_{i_l}\in
{\rm Mor}(X_{i_l},Y_1)$ such that $\phi=\sum_l q_{i_l}\circ p_{i_l}$.
Now $\phi$ is epi and we have $f_1\circ q_{i_l}=f_2\circ q_{i_l}$
by the above remark. Hence $f_1\circ\phi=f_2\circ\phi$
and by this $f_1=f_2$: $F$ is faithful.

$F$ satisfies $F(Y^\ast)\iso F(Y)^\ast$:
\begin{eqnarray*}
 F(Y^\ast)=\bigoplus_{X\in\nabla_0}
 {\rm Mor}(X,Y^\ast)\otimes F(X)
          \iso\bigoplus_{X\in\nabla_0}{\rm Mor}
          (X^\ast,Y^\ast)\otimes F(X^\ast)\\
          \iso\bigoplus_{X\in\nabla_0}{\rm Mor}
          (X,Y)^\ast\otimes F(X^\ast)
       \iso\bigoplus_{X\in\nabla_0}{\rm Mor}
       (X,Y)^\ast\otimes F(X)^\ast =F(Y)^\ast
  \end{eqnarray*}
Functoriality of $\ast$ is used in the third step and
the fourth step uses the fact that
$F(X)$ and $F(X^\ast)$ are vector spaces of equal dimension.

For every pair of  irreducible objects $X_1,X_2\in\nabla_0$
we choose an arbitrary (epi/iso)morphism

\[C_{X_1,X_2}:F(X_1)\otimes F(X_2)\pfeil F(X_1\tprod X_2)=
\bigoplus_{X\in\nabla_0}{\rm Mor}(X,X_1\tprod X_2)\otimes F(X)\]

$c$ is defined as extension of $C$:
\begin{eqnarray*}
c_{Y_1,Y_2}:F(Y_1)\otimes F(Y_2)\pfeil F(Y_1\tprod Y_2)\\
c_{Y_1,Y_2}:\left(\bigoplus_{X_1\in\nabla_0}{\rm Mor}(X_1,Y_1)
\otimes F(X_1)\right)
\otimes
\left(\bigoplus_{X_2\in\nabla_0}{\rm Mor}(X_2,Y_2)
\otimes F(X_2)\right)
\\ \pfeil
\bigoplus_{X\in\nabla_0}{\rm Mor}(X,Y_1\tprod Y_2)\otimes F(X)\\
c_{Y_1,Y_2}:=\bigoplus_{X_1,X_2\in\nabla_0}
(\Gamma\otimes id)\circ C_{X_1,X_2}\circ\tau_{2,3}\\
\Gamma:{\rm Mor}(X_1,Y_1)\otimes{\rm Mor}(X_2,Y_2)
\otimes{\rm Mor}(X,X_1\tprod X_2)\pfeil {\rm Mor}(X,Y_1\tprod Y_2)\\
\Gamma(f_1\otimes f_2\otimes g):=(f_1\tprod f_2)\circ g
\end{eqnarray*}
\end{bew}


Proposition \ref{funktorkon} reduces the problem of finding a functor
to finding a dimension function.
This is always possible:
\begin{satz} Let $\oalg$ be the observable algebra
of a QFT.
On $\kat:=[{\rm Rep}(\oalg)]$ there exist always weak dimension
functions. An example is
\begin{equation}
D_1(1):=1\qquad D_1(X):={\rm dim}\bigoplus_{Y,Z\in\nabla_0}
{\rm Mor}(Y\tprod X,Z)=\sum_{i,j}N_{X,i}^j
\end{equation}
In the algebraic formulation of QFT
we have another possibility.
\begin{equation}
 D_2(\rho):={\rm dim(span}\{(\rho_I\rho,\rho_J)\mid
 \rho_I,\rho_J\in\nabla_0\})
\end{equation}

A third possibility was found by Schomerus \cite{schomerus1}:
\begin{equation}
D_3(1):=1\qquad D_3(\rho):=C\qquad C:=max_{I,J\neq0}\sum_K N^K_{I,J}
\end{equation}
\end{satz}
\begin{bew}
\begin{eqnarray*}
D_1(X)D_1(Y)=\left(\sum_{s,r}N_{X,s}^r\right)\left(\sum_{S,R}N_{Y,S}^R\right)
=\sum_{s,r,S,R}N_{X,s}^rN_{Y,S}^R\geq\\
\sum_{K,N,M}N_{X,N}^KN_{Y,K}^M=\sum_{K,N,M}N_{X,Y}^KN_{K,N}^M=D_1(X\tprod Y)
\end{eqnarray*}
\end{bew}

\subsection{Examples of Quasi Tensor Categories}  \label{ex}

\subsubsection{Representation Categories of
quasitriangular ((weak) quasi) Hopf Algebras}

Representations of a Hopf algebra $H$ form  a
monoidal category ${\rm Rep}(H)$. The objects are the
representations and the morphisms are the
intertwiners between them. It is braided if $H$ is quasitriangular.
The braid isomorphism is naturally given by
\begin{equation}\label{zopfhopf}\Psi(v_1\tprod v_2):=\tau\circ
  (\varrho^1\otimes\varrho^2)(R)(v_1\tprod v_2)
\end{equation}
For triangular Hopf algebras this is  a symmetric tensor category, while
for QTHA it is a braided one.

The subcategory ${\rm Rep}(H)^{fd}$
of finite dimensional representations is rigid
thanks to the conjugated representation.

\subsubsection{Representation Categories of Observable Algebras}\label{aqfttk}

The category of local charge representations
of the algebra of observables $\oalg$
in algebraic quantum field theory is ideal: It possesses all
the properties defined in the beginning of this chapter.

\begin{eqnarray}
{\rm Obj}(\kat)&=&\nabla \qquad\mbox{proper morphisms$\iso$representations}\\
{\rm Mor}(\rho_1,\rho_2)&=&(\rho_2,\rho_1)\qquad\mbox{ intertwiner}\\
\tprod&=&\schlange{\times}\qquad\rho_1\schlange{\times}\rho_2:=
\rho_1\circ\rho_2\qquad
1=\rho_0=id\\
T_1\circ T_2&=&T_1T_2\qquad
T^\dagger=T^\ast\qquad \rho^\ast=\quer{\rho}\\
\Psi\iso\sigma_\epsilon\\
ev_X&\iso&\schlange{R}^\ast:=R^\ast d_\rho^{1/2}\label{ttid007}\qquad
coev_X\iso\schlange{\quer{R}}:=
   \epsilon_{\quer{\rho},\rho}\schlange{R}\label{ttid008}
\end{eqnarray}

\begin{bem} In general we use the term quantum field theory (QFT) in a
weaker sense that includes (among others) also CQFT (for example in the
axiomatic framework of \cite{ffk2}). We assume that one has a distinguished
algebra $\oalg$ (observable algebra, chiral Algebra which has direct sums,
tensor products, involution and charge conjugation (rigid)).
\end{bem}

\subsubsection{Ultra weak quasi Hopf algebras}

Is there some kind of algebra generalizing the ((weak) quasi) quantum groups
and observable algebras? We believe that ultra weak quasi quantum groups
as introduced in \cite{haring} may provide an answer.

\begin{de}[Ultra weak quasi Hopf algebra]
An {\em\bf $A$-ultra weak quasi Hopf algabra} $H$
($A$ an unital algebra)
is a $A$ bialgebra $H$
(left and right multiplication are denoted by
$\mu_l:A\otimes H\pfeil H, \mu_r:H\otimes A\pfeil H$)
and algebra morphisms $\eta:A\pfeil H,\epsilon:H\pfeil A$ such that
all axioms of a weak quasi Hopf algebra are fulfilled with the exception of
unit/counit properties which are replaced by:
\begin{equation}
\mu_l(\epsilon\otimes id)\Delta=\mu_r(id\otimes \epsilon)\Delta=id_H\qquad
m(id\otimes\eta)=\mu_r\qquad m(\eta\otimes id)=\mu_l
\end{equation}
\end{de}

\section{Reconstruction Theorems}

Historically the first reconstruction theorem was the famous
Tannaka-Krein theorem: Given a symmetric tensor category and a
functor to ${\underline\rm Vec}$ there is a group with the given category as
representation category. Majid has proved reconstruction theorems for
quasitriangular Hopf algebras and quasi Hopf algebras.
A reconstruction theorem for weak quasi Hopf algebras was suggested by
Kerler without a proof. However he demands some kind of symmetric-weak
tensor functor which cannot exist according to a simple argument by
Kenrik Kratz \cite{kratz}.

The {\em\bf forgetful functor}
$V:{\rm Rep}(H)\pfeil\underline{{\rm Vec}}$
assigns to each representation the underlying
vector space.

\subsection{Majid's Reconstruction Theorem}\label{mj}

\begin{theo}[Generalized Majid's first reconstruction theorem]
\label{majidrt1}
\footnote{The generalization to weak quasi Hopf algebras was suggested
by Kerler without proof and with wrong assumptions.
The correct formulation and the proof are belived to be new as are
the ultra weak case, the involution and the proof of injectivity and
surjectivity of $G$.}
Let $\kat$ be a rigid braided tensor category and
$F:\kat\pfeil\underline{{\rm Vec}}$
a monoidal functor. Then there is a maximal algebra
$(H,R)=H(\kat,F)$, unique upto isomorphism, and an induced  functor
$G:\kat\pfeil {\rm Rep}(H)$, such that
$\kat\stackrel{G}{\pfeil}{\rm Rep}(H)\stackrel{V}{\pfeil}
\underline{{\rm Vec}}$
composes to $F$.
$G$ maps inequivalent objects to inequivalent representations.
$G$ is always full and $G$ is faithful iff $F$ is faithful.
$G$ is surjective as a map of objects of irredundant categories
if $\kat$ is semisimple and Schur.\footnote{Hence
in the case of a faithful functor and a semisimple Schur category we have
$[\kat]\iso[{\rm Rep}(H)]$ and therefore
$\kat$ and ${\rm Rep}(H)$ are equivalent categories.}
${\rm Rep}(H)$ is rigid if $F$ is faithful and it is $C^\ast$ if $\kat$ is so.
The structure matrices (see section \ref{peq}) coincide.
The structure of $H$ is determined by $F$:

$F$ is tensor functor \quad$\Longrightarrow$\quad
$H$ is quasitriangular Hopf algebra

$F$ is quasi tensor functor \quad$\Longrightarrow$\quad
$H$ is quasitriangular quasi Hopf algebra

$F$ is weak quasi tensor functor \quad$\Longrightarrow$\quad
$H$ is quasitriangular weak quasi Hopf algebra

$F$ is ultra weak quasi tensor functor \quad$\Longrightarrow$\quad
$H$ is quasitr. ultra weak quasi Hopf algebra

If $\underline{Vec}$ is replaced by the category of finite dimensional
Hilbert spaces $H$ has an involution $\ast:H\pfeil H$
such that
$(gh)^\ast=h^\ast g^\ast, \Delta(h^\ast)=\Delta(h)^\ast, \epsilon(h^\ast)
=\quer{\epsilon(h)}, S^{-1}(h^\ast)=S(h)^\ast$ with
$(g\otimes h)^\ast=g^\ast\otimes h^\ast$.

$H$ is explicitly given by $H=\{h:{\rm Obj}(\kat)\pfeil
{\rm End_{{\rm Vec}}}\mid
 h_X\in{\rm End}(F(X)), F(f)\circ h_X=h_Y\circ F(f)\quad X,Y\in{\rm Obj}(\kat)
 f\in{\rm Mor}(X,Y)\}$. The functions $h$  are called covariant.
\end{theo}
\begin{bew}
$H$ becomes a vector space by pointwise addition.
The multiplication is also defined pointwise:
$(hg)_X:=h_X\circ g_X\quad X\in{\rm Obj}(\kat), h,g\in H$. The unit is
$X\mapsto 1_X={\rm id}_{F(X)}$.	(The ultra weak case is handled at the
end of the proof.)

In $\underline{{\rm Vec}}$
the following relation holds\footnote{In the Hilbert space case
this is only an injection. We than have to project on the image of this
injection. Alternatively we may regard  ${\rm End}(F(X))=B(F(X))$ as
von Neumann algebra over the hilbert space $F(X)$ and take
the weak closure of the tensor product factors.}:
 ${\rm End}(F(X))\otimes{\rm End}(F(Y))\iso
{\rm End}(F(X)\otimes F(Y))$. $H\otimes H$ is given by
functions in two variables $X,Y$,
which map to ${\rm End}(F(X)\otimes F(Y))$.
The coproduct $\Delta:H\pfeil H\otimes H$ is defined by:
\begin{equation}\Delta(h)_{X,Y}:=
c_{X,Y}^{-1}\circ h_{X\tprod Y}\circ c_{X,Y}\end{equation}
This is compatible with multiplication:
\begin{eqnarray*}
(\Delta(h)\Delta(g))_{X,Y}=\Delta(h)_{X,Y}\Delta(g)_{X,Y}
=c^{-1}_{X,Y}h_{X\tprod Y}c_{X,Y}c_{X,Y}^{-1}g_{X\tprod Y}c_{X,Y}
=\\c^{-1}_{X,Y}h_{X\tprod Y}g_{X\tprod Y}c_{X,Y}=\Delta(hg)_{X,Y}
\end{eqnarray*}

The counit is $\epsilon:H\pfeil\CC,\epsilon(h):=h_1$.
\[((1\otimes\epsilon)\Delta(h))_X=\Delta(h)_{X,1}=
c_{X,1}^{-1}h_{X\tprod 1}c_{X,1}=h_{X\tprod 1}=h_X\]

The associator
 $\phi\in H\otimes H\otimes H$ is given by
$\phi_{X,Y,Z}:=(c_{X,Y}^{-1}\otimes 1)c^{-1}_{X\tprod Y,Z}
F(\Phi_{X,Y,Z})c_{X,Y\tprod Z}(1\otimes c_{Y,Z})$.
For tensor functors this is trivial because of (\ref{assod}).
For quasi tensor functors it is invertible.
\begin{eqnarray*}(\phi(1\otimes\Delta)\Delta(h))_{X,Y,Z}
&=&\phi_{X,Y,Z} (c_{X,Y\tprod Z}(1\otimes c_{Y,Z}))^{-1}h_{X\tprod(Y\tprod Z)}
   c_{X,Y\tprod Z}(1\tprod c_{Y,Z})\\
  &=&(c_{X,Y}^{-1}\otimes 1)c^{-1}_{X\tprod Y,Z}F(\Phi_{X,Y,Z})
     h_{X\tprod(Y\tprod Z)}c_{X,Y\tprod Z}(1\otimes c_{Y,Z})
     \\
  ((\Delta\otimes1)\Delta(h)\phi)_{X,Y,Z}&=&
   (c^{-1}_{X,Y}\otimes1)c^{-1}_{X\tprod Y,Z}h_{(X\tprod Y)\tprod Z}
     c_{X\tprod Y,Z}(c_{X,Y}\otimes1)\phi_{X,Y,Z}\\
   &=&
   (c^{-1}_{X,Y}\otimes1)c^{-1}_{X\tprod Y,Z}h_{(X\tprod Y)\tprod Z}
     F(\Phi_{X,Y,Z})c_{X,Y\tprod Z}(1\otimes c_{Y,Z})\\
\end{eqnarray*}
Both expressions are the same because of covariance:
"$F(\Phi)h=hF(\Phi)$"
This shows quasi coassociativity. For
tensor functors this reduces to coassociativity
and for weak quasi tensor functors  $\phi$ remains quasi invertible.

For the proof of
$(id\otimes id\otimes\Delta)(\phi)\cdot(\Delta\otimes id\otimes id)(\phi)=
(1\otimes\phi)(id\otimes\Delta\otimes id)(\phi)(\phi\otimes1)$.
we refer to Majid's original work \cite{majid2} or \cite{haring}.

$F$ is a functor between rigid braided tensor categories.
There are isomorphisms $d_X:F(X)^\ast\iso F(X^\ast)$ and
$d_X^\ast:F(X^\ast)^\ast\iso F(X)$. They are used in the definition
of the antipode:
\begin{equation} (Sh)_X:=d_X^\ast(h_{X^\ast})^\ast d_X^{\ast-1}
\end{equation}
We omit the proof of the antipode identity\footnote{The antipode identity
(or even a stronger identity) is considered to be essential
by Mack/Schomerus for consistent transformation laws of adjoint field
operators. We don't need this assumption in our approach.}
which may be found in \cite{haring}.

$H$ is quasitriangular by means of  $R\in H\otimes H$:
\begin{equation}
R_{X,Y}:=\Psi^{{\rm Vec}-1}_{F(X),F(Y)}c^{-1}_{Y,X}F(\Psi_{X,Y})c_{X,Y}
\end{equation}

$R$ relates the coproduct and the opposite coproduct:
\begin{eqnarray*}
(R\Delta(h)R^{-1})_{X,Y}&=&\Psi^{{\rm Vec}-1}_{F(X),F(Y)}c^{-1}_{Y,X}
F(\Psi_{X,Y})c_{X,Y}c_{X,Y}^{-1}h_{X\tprod Y}c_{X,Y}\\
&&c_{X,Y}^{-1}F(\Psi_{X,Y})^{-1}c_{Y,X}\Psi^{{\rm Vec}}_{F(X),F(Y)}\\
&=& \Psi^{{\rm Vec}-1}_{F(X),F(Y)}c^{-1}_{Y,X}F(\Psi_{X,Y})h_{X\tprod Y}
F(\Psi_{X,Y})^{-1}c_{Y,X}\Psi^{{\rm Vec}}_{F(X),F(Y)}\\
&=& \Psi^{{\rm Vec}-1}_{F(X),F(Y)}c^{-1}_{Y,X}h_{\Psi(X\tprod Y)}
c_{Y,X}\Psi^{{\rm Vec}}_{F(X),F(Y)}\\
&=&\Delta'(h)_{X,Y}
\end{eqnarray*}
For the proof of the other two quasitriangularity equations
we refer once more to \cite{majid1} and \cite{haring}.

In the weak case one has in addition to verify:
\begin{eqnarray}
\phi^{-1}\phi&=&(id\otimes\Delta)\Delta(1)\label{ff1}\\
\phi\phi^{-1}&=&(\Delta\otimes id)\Delta(1)\label{ff2}\\
RR^{-1}&=&\Delta'(1)\label{ff3}\\
R^{-1}R&=&\Delta(1)\label{ff4}\\
(id\otimes id\otimes\epsilon)(\phi)&=&(id\otimes\epsilon\otimes id)(\phi)
=(\epsilon\otimes id\otimes id)(\phi)=\Delta(1)\label{ff5}\\
\end{eqnarray}

This is easily done using $cc^{-1}=1, c^{-1}c\neq1$: For (\ref{ff5})
we calculate:
\[(id\otimes id\otimes\epsilon)(\phi)_{X,Y}=
(c^{-1}_{X,Y}\otimes 1)c^{-1}_{X\tprod Y,1}F(\Phi_{X,Y,1})c_{X,Y}
(1\otimes c_{Y,1})
=(c^{-1}_{X,Y})c_{X,Y}=\Delta(1)_{X,Y}\]
And for (\ref{ff4}):\[(R^{-1}R)_{X,Y}=
c_{X,Y}^{-1}F(\Psi_{X,Y}^{-1})c_{Y,X}\Psi^{{\rm Vec}}_{F(X),F(Y)}
\Psi^{{\rm Vec}-1}_{F(X),F(Y)}c^{-1}_{Y,X}F(\Psi_{X,Y})c_{X,Y}
=c_{X,Y}^{-1}c_{X,Y}=\Delta(1)_{X,Y}\]
Similarly one gets\footnote{This calculation was carried out by H. Kratz.}
 (\ref{ff2}):
\begin{eqnarray*}
& &\phi_{X,Y,Z}\circ \phi^{-1}_{X,Y,Z}\\
&=&(c_{X,Y}^{-1}\otimes id)c_{X\otimes Y,Z}^{-1}
F(\Phi_{X,Y,Z})c_{X,Y\otimes Z}(id\otimes c_{Y,Z})\\&&
(id\otimes c_{Y,Z}^{-1})c^{-1}_{X,Y\otimes Z}
F(\Phi_{X,Y,Z})^{-1}c_{X\otimes Y,Z}(c_{X,Y}\otimes id)\\
&=&(c_{X,Y}^{-1}\otimes id)c_{X\otimes Y,Z}^{-1}
c_{X\otimes Y,Z}(c_{X,Y}\otimes id)\\
&=&(c_{X,Y}^{-1}\otimes id)
\Delta(1)_{X\otimes Y,Z}(c_{X,Y}\otimes 1)\\
&=&((\Delta \otimes id) \Delta(1))_{X,Y,Z}.
\end{eqnarray*}
(\ref{ff1}) is proven in the same way, just as (\ref{ff3}).

The vector spaces  $F(X)$ are  representation spaces of $H$:
$\varrho_X(h).v:=h_X(v)\quad h\in H,v\in F(X)$.
This induces a functor $\kat\pfeil{\rm Rep}(H)$.
Morphisms $f\in{\rm Mor}(X,Y)$ are mapped to intertwiners $G(f)=F(f)$:
$G(f)\circ\varrho_X(h)=F(f)\circ h_X=h_Y\circ F(f)=\varrho_Y(h)\circ G(f)$.

The proof of surjectivity needs lemma \ref{lemfl}. It shows that for
semisimple
Schur categories $H$ is a direct sum of full matrix algebras $M_n(\CC)$.
Each of them has only one irrep. And so $H$ has no other irreducible
representations, because all representations have to reflect commutativity
of the summands and must therefore annihilate all summands but one.
Therefore $H$ has no more irreducible representations than $[\kat]$ has
irreducible objects.
Since $F$ and $G$ are linear this shows that $G$ is surjective
when $G$ is considered as  a map of objects of irredundant categories.

$G$ is full, because every morphism $T$ in ${\rm Rep}(H)$
($T\varrho^Y=\varrho^XT$)
is a constraint that can only exist if it is of the form $T=F(f)$.

The proof of injectivity is simpler: Assume $X,Y$ to be inequivalent
objects which are mapped to equivalent representations, i.e.
$F(X)=F(Y),\forall h\in H,
h_X=\varphi\circ h_Y\circ\varphi^{-1}$ with an isomorphism
$\varphi:F(X)\pfeil F(Y)=F(X)$. So the value of
$h$ on $X$ is determined uniquely
by its value on $Y$. This can be done by covariance only if
$\exists f\in{\rm Mor}(X,Y)\exists g\in{\rm Mor}(Y,X)$
such that $F(f)=\varphi,
F(g)=\varphi^{-1}$.
But then (by faithfulness) $f$ and $g$ are iso ($id_{F(Y)}=F(f)F(g)=F(fg)$;
because of faithfulness only $id_Y$ is mapped to $id_{F(Y)}$ and hence
$f=g^{-1}$) contracting our hypothesis.

Describe $\kat$ as in subsection \ref{peq}.
According to this presentation we have for $X,Y,Z\in\nabla_0$
matrices $\Omega$ that satisfy $\phi^i\CVO{X}{Y}{Z}\circ\Psi_{Y,X}=\sum_j
\Omega_{i,j}\phi^j\CVO{Y}{X}{Z}$. We apply $F$, multiply $c$ from the right,
introduce $1=cc^{-1}$ and use linearity of $F$ to get
$F(\phi^i\CVO{X}{Y}{Z})\circ c\circ c^{-1}\circ\Psi_{Y,X}\circ c=\sum_j
\Omega_{i,j}F(\phi^j\CVO{Y}{X}{Z})\circ c$.
This shows\footnote{Note that $F(\phi^i\CVO{X}{Y}{Z})\circ c$
form a base of morphisms in ${\rm Mor}(F(X)\otimes F(Y),F(Z))$.
They are linearly independent: Assume $\sum_i\alpha_i
F(\phi^i\CVO{X}{Y}{Z})\circ c$.
By surjectivity of $c$ and linearity of $F$ this implies
$0=F(\sum_i\alpha_i \phi^i\CVO{X}{Y}{Z}))$
and faithfulness of $F$ yields a contradiction.
Further they span the whole space since $G$ is full.}
that $\kat$ and ${\rm Rep}(H)$ have the same structure constants.

${\rm Rep}(H)$ is $c^\ast$ by $F(f)^\dagger=F(f^\dagger)$ and it
is rigid if $F$ is faithful:
To $\varrho_X$ there is the dual representation
$\varrho_{X^{\ast}}$ such that $\varrho_X\iso\varrho_{X^{\ast\ast}}$.
The internal Hom is ${\rm Hom}(X,Y)=X^\ast\tprod Y$. To proof this one
has to show
${\rm Mor}(\varrho_Z\tprod\varrho_X,\varrho_Y)\iso
 {\rm Mor}(\varrho_Z,\varrho_{X^\ast}\tprod\varrho_Y)$.
The isomorphism $\alpha$ is explicitly given by the follwing construction:
Take an intertwiner $T\in{\rm Mor}(\varrho_Z\tprod\varrho_X,\varrho_Y)$
\[\varrho_Y(h)T=T(\varrho_Z\tprod\varrho_X)(h)
\Leftrightarrow h_YT=Tc^{-1}h_{Z\tprod X}c
\Rightarrow h_Y T c^{-1}=Tc^{-1}h_{Z\tprod X}\]
This implies $Tc^{-1}=F(f)$ and if $F$ is faithful $f$ is unique.
$\kat$ is rigid $\Rightarrow\exists!g\in{\rm Mor}(Z,X^\ast\tprod Y)$
\[h_{X^\ast\tprod Y}F(g)=F(g)h_Z\Leftrightarrow
c(\varrho_{X^\ast}\tprod\varrho_Y)(h)c^{-1}F(g)=F(g)\varrho_Z(h)\]
\[\Leftrightarrow
c^{-1}c(\varrho_{X^\ast}\tprod\varrho_Y)(h)c^{-1}F(g)=c^{-1}F(g)\varrho_Z(h)
\]
$c^{-1}c$ is nothing but $\Delta(1)$ and can therefore be omited.
It is easy to see that $\alpha(T):=c^{-1}F(g)$ constructed this way is iso.

The involution is given by: $(h^\ast)_X:=(h_X)^\ast$.
Multiplicativity carries over from vector space endomoephisms.
$\Delta(h^\ast)=\Delta(h)^\ast$ can be proven by assuming the
$c_{X,Y}$ without loss of generality to be isometries:
$\Delta(h)^\ast_{X,Y}=(c_{X,Y}^{-1}\circ h_{X\tprod Y}\circ c_{X,Y})^\ast=
c_{X,Y}^{-1}\circ h_{X\tprod Y}^\ast \circ c_{X,Y}$.
If $d$ is unitary then
$S(h)^\ast=S^{-1}(h^\ast)$, because of:
\begin{eqnarray*}
S(S(h)^\ast)_X=d_X^\ast\circ((S(h)^\ast)_{X^\ast})^\ast\circ d_X^{\ast-1}
=d_X^\ast\circ((S(h)_{X^\ast}^\ast)^\ast\circ d_X^{\ast-1}
=d_X^\ast\circ(S(h)_{X^\ast})\circ d_X^{\ast-1}\\
=d_X^\ast\circ d^\ast_{X^\ast}\circ h_X^\ast\circ d_{X^\ast}^{\ast-1}
\circ d_X^{\ast-1}
=d_X^\ast\circ d_X\circ h_X^\ast\circ d_X^{-1}\circ d_X^{\ast-1}
=h_X^\ast=(h^\ast)_X
\end{eqnarray*}
The square of the antipode is in the unitary case:
$S(S(h))_X=d_X^\ast\circ(S(h)_{X^\ast})^\ast\circ d_X^{\ast-1}
=d_X^\ast\circ(d_{X^\ast}^\ast\circ h_X^\ast\circ
d_{X^\ast}^{\ast-1})^\ast\circ d_X^{\ast-1}
=d_X^\ast\circ d_{X^\ast}^{-1}\circ h_X\circ d_{X^\ast}\circ d_X^{\ast-1}
=d_X^\ast\circ d_X\circ h_X\circ d^\ast_X\circ d_X
=(u^\ast h u)_X$ with $u_X:=d_X^\ast\circ d_X$.

The representations of $H$ are unitary:
$\varrho(h^\ast)=\varrho(h)^\ast$.

Let's have a look at the ultra weak case.
$H$ becomes a ${\rm End}(F(1))$-ultra weak Quasi-Hopfalgebra
with the following bimodule actions:
\begin{equation}
\mu_l(a\otimes h)_X:=c_{1,X}\circ(a\otimes h)\circ c^{-1}_{1,X}\quad
 \mu_r(h\otimes a)_X:=c_{X,1}\circ(h\otimes a)\circ c^{-1}_{X,1}\quad
 a\in{\rm End}(F(1))
 \end{equation}
The definition of $\epsilon$ doesn't have to be changed but
the unit is now defined more general to be
\begin{equation}
\eta(a):=\mu_l(a\otimes 1)=\mu_r(1\otimes a)
\end{equation}
The counit property is fulfilled:
\[(\mu_l(\epsilon\otimes id)\Delta(h))_X=c_{1,X}c_{1,X}^{-1}\circ
h_{1\tprod X}\circ c_{1,X}c_{1,X}^{-1}=h_X\]
\end{bew}

\begin{bem}\label{rtbem1}\begin{enumerate}
\item The existence of a weak quasi tensor functor for the representation
categories of observable algebras in quantum field theories
was already proven,
so that we can now state our first main result:
{\em\bf Every quantum field theory posesses a (not uniquely determined)
weak quasi Hopf algebra as possible gauge symmetry algebra.}
\item Let $F$ be as in proposition \ref{funktorkon}.
The choice of the arbitrary epimorphisms
$C_{X,Y}$ has no impact on $H(\kat,F)$.
Suppose that $\schlange{C}_{X,Y}$ is another choice and
denote by $\schlange{F}$ the functor constructed this way.
$\schlange{c}_{X,Y}:\schlange{F}(X)\otimes\schlange{F}(Y)\pfeil
\schlange{F}(X\tprod Y)$. Since $c,\schlange{c}$ are epi there are
isomorphisms $\phi_{X,Y}$ such that
$\schlange{c}_{X,Y}=\phi_{X,Y}\circ c_{X,Y}$.
Now take as an example $\schlange{h}\in H(\kat,\schlange{F})$
and calculate its coproduct:
$\Delta(\schlange{h})_{X,Y}=\schlange{c}^{-1}_{X,Y}\schlange{h}_{X\tprod Y}
\schlange{c}_{X,Y}=c^{-1}_{X,Y}\phi^{-1}_{X,Y}\schlange{h}_{X\tprod Y}
\phi_{X,Y}c_{X,Y}$.
We see that
$\schlange{h}\mapsto \phi^{-1}\circ\schlange{h}\circ\phi$
is an isomorphism between the 'two-point-evaluation' of
functions in
$H(\kat,\schlange{F})$  and $H(\kat,F)$.
Similar reasoning applies also to the case of $n-point-evaluations$.
\item The weakening procedure of Mack and Schomerus that associates to
a quasitriangular Hopf algebra $H$ a quasitriangular weak quasi Hopf algebra
${\cal W}(H)$ that isn't plagued by unphysical representations
can now be understood in the following manner:
${\cal W}(H)=H({\cal S}({\rm Rep}(H)),V)$.
\item Let $H$ be a finite dimensional ((weak) quasi) Hopf algebra.
There is a natural
injection\footnote{This is injective because every finite dimensional algebra
posesses a faithful representation.}
$i:H\pfeil H({\rm Rep}(H),V)$ given by $i(h)_X:=\varrho^X(h)$.
If $H$ is semisimple this is also surjective by Wedderburns structure theorem.
\end{enumerate}
\end{bem}

\subsubsection{Constructive Reconstruction}   \label{konreka}

Let $F$ be a (weak) quasi tensor functor constructed according to
proposition \ref{funktorkon}.
We want to explore the structure of the reconstructed (weak) quasi Hopf
algebra in more detail and want to carry out the reconstruction
for the Ising model.

\begin{lemma} \label{lemfl} Let $\kat$ be a
semisimple, rigid
braided tensor Schur category and $F$ as in proposition \ref{funktorkon}.
The covariant functions building up $H$ are uniquely determined by their
values on one object out of every irreducible equivalence class.
On those they may be arbitrary while
on the other members of
the equivalence class the value is  determined by covariance.
\end{lemma}
\begin{bew}
We restrict ourselves to the case of an
object consisting of two irreducible objects $Y=X_1\schlange{\oplus}X_2$.
The general case follows in the same way.

Let $f_i\in{\rm Mor}(X_i,Y),i=1,2$. By covariance we have
$F(f_i)h_{X_i}=h_YF(f_i)$ and $F(Y)=\KK f_1\otimes F(X_1)\bigoplus
\KK f_2\otimes F(X_2)$.
$F(f_i):F(X_i)\pfeil\KK f_1\otimes F(X_1)\bigoplus\KK f_2\otimes F(X_2),
x_i\mapsto f_i\otimes x_i$.
The inverse is
$F(f_i)^{-1}:f_1\otimes x_1\bigoplus f_2\otimes x_2\mapsto x_i$.
We have $F(f_i)^{-1}F(f_i)=id_{F(X_i)}$ and
$F(f_i)F(f_i)^{-1}=pr_i$.
Right multiplication of the covariance equation by
$F(f_i)^{-1}$ yields
$F(f_i)h_{X_i}=h_Ypr_i$.
This shows $h_Y$ to be determined uniquely by $h_{X_i}$.

On irreducible objects in different equivalence classes
covariant functions can take arbirary values because
there are no morphisms (and hence no restrictions)
between inequivalent irreducible objects.

If $g\in{\rm Mor}(Z_1,Z_2)$ is an isomorphism
between  irreducible, equivalent objects. Then
$F(g)$ is also iso and covariance determines $h_{Z_2}$ by the value $h_{Z_1}$.
\end{bew}

\begin{bsp}[Ising model]
Let $\kat:={\rm Rep}({\rm Vir}(c=1/2))$ be the
representation category of the Ising model. $\kat$ contains three
irreducible equivalence classes: $[0],[1/2],[1]$,
the equivalence classes of {\rm Vir}-representations with highest
weight 0,1/16,1/2.
Let $F$ be the weak quasi tensor functor constructed  as
in proposition \ref{funktorkon} by means of the weak dimension
function $D([I]):=2I+1$.
$H:=H(\kat,F)$ denotes the reconstructed weak quasi Hopf algebra.
The elements of $H$ have to be defined according to the lemma
on the irreducible objects. On those they may be arbitrary.

We define generators $e,f,h\in H$ to be the representation matrices of
${\cal U}_q(sl_2),q=-i$.
\begin{eqnarray*}
e_0:=h_0:=f_0:=0\in{\rm End}(\CC)=\CC\\
h_{1/2}:=\left(\begin{array}{cc}1&0\\0&-1\end{array}\right)\quad
e_{1/2}:=\left(\begin{array}{cc}0&1\\0&0\end{array}\right)\quad
f_{1/2}:=\left(\begin{array}{cc}0&0\\1&0\end{array}\right)\\
h_1:=\left(\begin{array}{ccc}-2&0&0\\0&0&0\\0&0&2\end{array}\right)\quad
e_1:=\left(\begin{array}{ccc}0&0&0\\a&0&0\\0&a&0\end{array}\right)\quad
f_1:=\left(\begin{array}{ccc}0&a&0\\0&0&a\\0&0&0\end{array}\right)\\
a:=[2]_q
\end{eqnarray*}
\begin{enumerate}
\item $e,f,h$ generates $H$: The threefold products suffice two
generate the full endomorphism algebras\footnote{This was
checked with a small Mathematica program which calculates the
dimension of the algebra. It is available from the
author upon request.}.
\item Beeing representation matrices of ${\cal U}_q(sl_2)$ the
generators fulfill the defining relations of this Hopf algebra:
$H$ is as an algebra a quotient of ${\cal U}_q(sl_2),q=-i$.
An expression $G(e,f,h)\in H$ vanishes iff it vanishes in all three
physical irreps of ${\cal U}_q(sl_2),q=-i$
So we have  $H={\cal U}_q(sl_2)/{\cal I}$, with ${\cal I}$
the ideal beeing annihilated by all physical representations
of ${\cal U}_q(sl_2)$. This shows that $H$  as an algebra
coincides with the algebra constructed by Mack/Schomerus.
\item  $\Delta(1)=c^{-1}\circ c$ is the projector on physical
subrepresentations. This shows $\Delta, R$ to be the same.
$S=\ast$ however diverges from the Mack/Schomerus antipode.
\end{enumerate}
\end{bsp}

\section{Gauge and Quantum Symmetry}  \label{gauge}

The picture of gauge theories as developed by Doplicher, Haag and Roberts
looks like this:
There is a field algebra $\falg$ and a gauge algebra
$\eich$ acting on it. The algebra of observables
$\oalg\subset\falg$ consists of those fields that are $\eich$ invariant.
The irreducible representations $\pi_I, I\in{\cal I}$ of $\oalg$ and those of
$\eich$,
$\varrho^I, I\in{\cal I}$ are 1-1 correlated.
The total Hilbert space takes the form
\begin{equation}\label{hilbertr}
 \hilb=\bigoplus_{I\in{\cal I}} \hilb_I\otimes V^I \end{equation}
Here $\hilb_I, V^I$ are the representation spaces of $\pi_I, \varrho^I$.

The general aim is to construct all these out of the algebra of
observables since this is the only part which can be determined by
observation.
A first step in this direction was undertaken in the algebraic formulation
of QFT (AQFT for short) where the reduced field bundle $\falg_r$ was
introduced  as a replacement for the field algebra when the symmetry
is not known.
In $\falg_r$ vertex operators can be defined\footnote{See \cite{haring}
for details.}.
It is widely belived that two dimensional conformal QFT are tractable
in the framework of AQFT. In both theories the vertex operators are
intertwining operators between irreps and products of irreps.
They satisfy braid and fusion idenitities which encode the structure
of the $\oalg$ representation category and show it to be a braided braided
tensor
category.

To act as a symmetry an algebra has to have a representation category
that coincides with that of $\oalg$:
\\[0.5ex]
\begin{tabular}{rcl}
algebra of observables $\oalg$ & $\longleftrightarrow $ & gauge algebra
$\eich$\\[0.4ex]
Braiding of representations $\Psi,\epsilon$&$
\longleftrightarrow$& $R$ element\\
CVO & $\longleftrightarrow $ &  Clebsch Gordan  coefficient\\
fusion matrix $F$ & $\longleftrightarrow $ & 6-j symbols\\
braid matrix $B$ & $\longleftrightarrow $ & ---
\end{tabular}

The following list collects all requierments a gauge algebra has to fulfill.
\begin{enumerate}
\item {\em\bf Products of representations:}
       Produkts of $\eich$-covariant field operators will transform
       under some tensor product representation. Therefore $\eich$
       must have a coproduct (or somthing similar) to allow products of
       representations. Neither coassociativity nor cocommutativity are
       requiered a priori.
\item {\em\bf Structure of the representation category:}
       We need to associate a $\eich$ representaion to every equivalence
       class of $\oalg$ representations. The reduction of tensor products
       has to be the
       same.
       Put in a mathematical language: We need a
       tensor functor $G:{\rm Rep}(\oalg)\pfeil{\rm Rep}(\eich)$
       that maps inequivalent irreps to inequivalent irreps and
       preserves the structure constants.
\item {\em\bf Invariance of vacuum:} The vacuum has to be
invariant\footnote{This
requirement rules out ultraweak quasi Hopf algebras as symmetry algebras.
They would turn the gauge symmetry into a broken symmetry.}, i.e.
       it must transform under a one dimensional representation:
       \begin{equation}\varrho(a)\vak=\vak\epsilon(a) \end{equation}
\item {\em\bf Unitarity:}
       To maintain a quantum mechanical interpretation we need a Hilbert space
       where the gauge algebra acts unitarily:
       \begin{equation} \varrho(a^\ast)=\varrho(a)^\ast \quad
       \Delta(a^\ast)=\Delta(a)^\ast\quad
       \epsilon(a^\ast)=\epsilon(a)^\ast\end{equation}
\end{enumerate}

The weak quasi Hopf algebra reconstructed in the last chapter fulfills
this requirement. Our second step is now to build a gauge covariant
field algebra.

{\em\bf Preliminaries:}
We choose bases $e^I_i. i=1\ldots{\rm dim}(V^I)$ in the
representation spaces $V^I$ of the $\eich$ representation $\varrho^I$
and in the morphism spaces:
$\chi\CVO{C}{S}{R}\in{\rm Mor}(\varrho^C\otimes\varrho^S,\varrho^R)$.
The $\chi$ are analogous to  Clebsch Gordan coefficients:
\begin{equation}
\chi\CVO{C}{S}{R}(e^C_c\otimes e^S_s)=\sum_r\clebsch{C}{S}{R}{c}{s}{r}e^R_r
\end{equation}
Sine the representation categories of $\eich$ and $\oalg$ are equivalent
the braiding and fusion in $\oalg$ carry over from braiding and fusion
of vertex operators $\phi(e)$ of $\oalg$.
\begin{eqnarray}
\chi(e_2)(id\otimes\chi(e_2))&=&
\sum_{e'_1,e'_2}B^\pm_{e_1,e_2;e'_2,e'_1}\chi(e'_1)(id\otimes\chi(e'_2))\\
\chi(e_2)(id\otimes\chi(e_2))&=&
\sum_{e,f}F_{e_1,e_2;f,e}\chi(e)(\chi(f)\otimes id)\\
\phi(e_2)(id\otimes\phi(e_2))&=&
\sum_{e'_1,e'_2}B^\pm_{e_1,e_2;e'_2,e'_1}\phi(e'_1)(id\otimes\phi(e'_2))\\
\phi(e_2)(id\otimes\phi(e_2))&=&
\sum_{e,f}F_{e_1,e_2;f,e}\phi(e)(\phi(f)\otimes id)\\
\end{eqnarray}
We use frequently multiindices $e=^\alpha\CVO{C}{S}{R}$ and call
$c(e)=C$ the charge, $s(e)=S$ the source and $r(e)=R$ the
range of $e$ or $\phi(e)$. $\alpha=1\ldots{\rm dim}({\rm Mor}
(C\otimes S,R))$

The same argument shows that the adjoint coefficients coincide.
\begin{eqnarray}
\phi(e)(h_C\otimes\cdot)^\ast&=&\sum_{e^\ast}\schlange{\eta}_{e,e^\ast}
\phi(e^\ast)(\schlange{h_C}\otimes\cdot)\label{konjform}\\
\chi(e)(v^C\otimes\cdot)^\ast&=&\sum_{e^\ast}\schlange{\eta}_{e,e^\ast}
\chi(e^\ast)(\schlange{v^C}\otimes\cdot)
\end{eqnarray}
Where $\schlange{\eta}_{e,e^\ast}$ is upto a normalization
the usual  $\eta$ matrix of AQFT.
The vectors $\schlange{h_C}\in\hilb_{C^\ast},
\schlange{v^C}\in V^{C^\ast}$ are determined uniquely but
they are irrelevant for our discussion.

Matrix elements of $R\in \eich\otimes\eich$ and
$\phi\in \eich\otimes\eich\otimes\eich$
are written according to the following example:
$\phi^{C_1,C_2,C_3,c'_1,c'_2,c'_3}_{c_1,c_2,c_3}=
(\varrho^{C_1,c'_1}_{c_1}\otimes
 \varrho^{C_2,c'_2}_{c_2}\otimes
 \varrho^{C_3,c'_3}_{c_3})(\phi)$

The action on basis vectors is
\begin{equation}
\sum_{c'_3,c'_2,c'_1}\phi^{C_3,C _2,C_1;c'_3,c'_2,c'_1}_{c_3,c_2,c_1}
e^{C_3}_{c'_3}\otimes(e^{C_2}_{c'_2}\otimes e^{C_1}_{c'_1})=
\phi((e^{C_3}_{c_3}\otimes e^{C_2}_{c_2})\otimes e^{C_1}_{c_1}
\end{equation}

\subsection{Vertex SOS Transformation}

The following concatenation of vertex operators can be carried out
in two ways:

\sbegin{equation}\label{trafo1}\chi\CVO{C_2}{Q}{R}(id\otimes
\chi\CVO{C_1}{S}{Q})
(e^{C_2}_{c_2}\otimes (e^{C_1}_{c_1}\otimes e^S_s))
=\sum_{q,r}\clebsch{C_1}{S}{Q}{c_1}{s}{q}\clebsch{C_2}{Q}{R}{c_2}{q}{r}e^R_r
\send{equation}
\sbegin{eqnarray}
\lefteqn{\chi\CVO{C_2}{Q}{R}(id\otimes\chi\CVO{C_1}{S}{Q})
(e^{C_2}_{c_2}\otimes(e^{C_1}_{c_1}\otimes e^S_s))=}\nonumber\\
&=&\sum_{\schlange{c_1},\schlange{c_2},\schlange{s}}
\phi^{-1;C_2,C_1,S,\schlange{c_2},\schlange{c_1},\schlange{s}}_{c_2,c_1,s}
\chi\CVO{C_2}{Q}{R}(id\otimes\chi\CVO{C_1}{S}{Q})
((e^{C_2}_{\schlange{c_2}}\otimes e^{C_1}_{\schlange{c_1}})
\otimes e^S_{\schlange{s}})= \nonumber\\
&=&\sum_{\schlange{c_1},\schlange{c_2},\schlange{s}}
\phi^{-1;C_2,C_1,S,\schlange{c_2},\schlange{c_1},\schlange{s}}_{c_2,c_1,s}
\sum_{c'_1,c'_2}\rmat^{-1C_2,C_1;c'_2,c'_1}_{\schlange{c_2},\schlange{c_1}}
  \nonumber\\&&
  \sum_{P,p}B^+
  \chi\CVO{C_1}{P}{R}(id\otimes\chi\CVO{C_2}{S}{P})
  ((e^{C_1}_{c'_1}\otimes e^{C_2}_{c'_2})\otimes e^S_{\schlange{s}})=
  \nonumber\\
&=&\sum_{\schlange{c_1},\schlange{c_2},\schlange{s}}
\phi^{-1;C_2,C_1,S,\schlange{c_2},\schlange{c_1},\schlange{s}}_{c_2,c_1,s}
\sum_{c'_1,c'_2}\rmat^{-1C_2,C_1;c'_2,c'_1}_{\schlange{c_2},\schlange{c_1}}
  \sum_{P,p}B^+
  \chi\CVO{C_1}{P}{R}(id\otimes\chi\CVO{C_2}{S}{P})\nonumber\\
 &&\sum_{\schlange{c_1}',\schlange{c_2}',\schlange{s}'}
 \phi^{C_1,C_2,S,\schlange{c_1}',\schlange{c_2}',\schlange{s}'}
 _{c'_1,c'_2,\schlange{s}}
  (e^{C_1}_{\schlange{c'}_1}\otimes(e^{C_2}_{\schlange{c'}_2}
  \otimes e^S_{\schlange{s}'}))=\nonumber\\
&=&\sum_{\schlange{c_1},\schlange{c_2},\schlange{s}}
\phi^{-1;C_2,C_1,S,\schlange{c_2},\schlange{c_1},\schlange{s}}_{c_2,c_1,s}
\sum_{c'_1,c'_2}\rmat^{-1C_2,C_1;c'_2,c'_1}_{\schlange{c_2},\schlange{c_1}}
\sum_{\schlange{c_1}',\schlange{c_2}',\schlange{s}'}
 \phi^{C_1,C_2,S,\schlange{c_1}',\schlange{c_2}',\schlange{s}'}_
 {c'_1,c'_2,\schlange{s}}\nonumber\\
&&  \sum_{P,p}B^+_{\CVO{C_1}{S}{Q},\CVO{C_2}{Q}{R};\CVO{C_2}{S}{P},
\CVO{C_1}{P}{R}}\sum_r\clebsch{C_2}{S}{P}{\schlange{c'}_2}{\schlange{s}'}{p}
   \clebsch{C_1}{P}{R}{\schlange{c}'_1}{p}{r}e^R_r
\label{trafo2}
\send{eqnarray}

$B^+:=B^+_{\CVO{C_1}{S}{Q},\CVO{C_2}{Q}{R};\CVO{C_2}{S}{P},
\CVO{C_1}{P}{R}}$.

(\ref{trafo1})=(\ref{trafo2}) implies:
\begin{satz}[vertex SOS transformation]
\sbegin{eqnarray}
\sum_q\clebsch{C_1}{S}{Q}{c_1}{s}{q}
  \clebsch{C_2}{Q}{R}{c_2}{q}{r}
  =\sum_{P,p,c'_1,c'_2,\schlange{c_1},\schlange{c_2},\schlange{s}}
  \rmat^{-1,C_2,C_1;c'_2,c'_1}_{\schlange{c_2},\schlange{c_1}}
\phi^{-1;C_2,C_1,S,\schlange{c_1},\schlange{c_2},\schlange{s}}_{c_2,c_1,s}
\nonumber\\
B^+_{\CVO{C_1}{S}{Q},\CVO{C_2}{Q}{R};\CVO{C_2}{S}{P},
\CVO{C_1}{P}{R}}
\sum_{\schlange{c_1}',\schlange{c_2}',\schlange{s}'}
 \phi^{C_1,C_2,S,\schlange{c_1}',\schlange{c_2}',\schlange{s}'}_
 {c'_1,c'_2,\schlange{s}}
  \clebsch{C_2}{S}{P}{\schlange{c'}_2}{\schlange{s}'}{p}
   \clebsch{C_1}{P}{R}{\schlange{c}'_1}{p}{r}
\send{eqnarray}
If $\eich$ is coassociative this reduces to
\sbegin{equation}
\sum_q\sum_{c_1,c_2}\clebsch{C_1}{S}{Q}{c_1}{s}{q}
  \clebsch{C_2}{Q}{R}{c_2}{q}{r}
  \rmat^{C_2,C_1;c_2,c_1}_{c''_2,c''_1} = \sum_{c_1,c_2}
   \sum_{P,p}B^+\clebsch{C_2}{S}{P}{c''_2}{s}{p}
   \clebsch{C_1}{P}{R}{c''_1}{p}{r}
\send{equation}
\end{satz}
This relation is called {\em\bf vertex SOS transformation}\footnote{It
was first postulated in \cite{ffk2} for Hopfalgebras.}.
Obviously one can transfer the inversion from
${\cal R}$ to $B$.

The vertex SOS transformation is seen from the point of $\eich$:
It relates braiding of $\eich$ representations ($\rmat$) via
$\eich$ Clebsch Gordan coefficients to the braiding of
$\eich$ vertex operators.
Since the representation categories of $\eich$ and $\oalg$ are equivalent
it is clear that there must exist also a vertex SOS transformation
for the $\oalg$ quantities.
In AQFT this relation is well known:
\begin{equation}
\sum_{e_1,e_2}B^\pm_{e_1,e_2;e'_2,e'_1}T_{e_1}T_{e_2}=\rho_\alpha(\epsilon)
T_{e'_2}T_{e'_1}                                        \label{btt}
\end{equation}

\subsection{Field Algebra $\falg_{u1}$}

In this section we will construct a covariant field algebra.
We have to make some technical assumptions which we mark
by TA.

$\falg_{u1}$ operates on (\ref{hilbertr}):
$\hilb=\bigoplus_I\hilb_I\otimes V^I$. $P^\oalg_I:\hilb\pfeil\hilb_I,
P^\eich_I:\hilb\pfeil V^I$  are the natural projections.
$\falg_{u1}$ is generated by
$\eich$, $\oalg$ and  special intertwiners.
We have natural embeddings $i_\oalg:\oalg\pfeil\falg_{u1}$ and
$i_\eich:\eich\pfeil\falg_{u1}$ operating on $h_I\otimes
v^I\in\hilb_I\otimes V^I$
by $i_\oalg(A)(h_I\otimes v^I):=\pi_I(A)h_I\otimes v^I$ and
$i_\eich(g)(h_I\otimes v^I):=h_I\otimes \varrho^I(g)v^I$.
$i_\eich$ and $i_\oalg$ commute.
Intertwiners between sectors are:

\begin{de} $\falg_{u1}$ is generated by
$i_\oalg(\oalg),i_\eich(\eich)$ and intertwiners:
\begin{eqnarray}
\lefteqn{\Psi^C(h_C\otimes v^C):\hilb\pfeil\hilb}
\end{eqnarray}\begin{eqnarray}
\Psi^C(h_C\otimes v^C):=\sum_{e,f,c(e)=c(f)=C} D_{e,f}
\phi(e)(h_C\otimes \cdot)\otimes\chi(f)(v^C\otimes \cdot)\label{psidef}\\
 v^C\in V^C, h_C\in\hilb_C
\end{eqnarray}

$D:W\otimes W\pfeil\CC,(e,f)\mapsto D_{e,f}\in\CC$ has to
fulfill the following relations which are needeed to proof
braid relations in $\falg_{u1}$.
\begin{eqnarray}            \label{ddef}
\sum_{r(e_1),r(f_1)} B^+_{e_1,e_2;e'_2,e'_1}B^-_{f_1,f_2;f'_2,f'_1}
D_{e_1,f_1}D_{e_2,f_2}
=D_{e'_1,f'_1}D_{e'_2,f'_2}\quad\mbox{(TA:DB)}\quad\\
c(e_1)=c(f_1),c(e_2)=c(f_2),s(e_2)=r(e_1),
s(e'_1)=r(e'_2),s(f_2)=r(f_1),s(f'_1)=r(f'_2)
\end{eqnarray}
For the proof of fusion rules we need further
\begin{eqnarray}            \label{ddef2}
\sum_{r(e_1),r(f_1)} F_{e_1,e_2;f,e}F_{f_1,f_2;\schlange{f},\schlange{e}}
D_{e_1,f_1}D_{e_2,f_2}
=D_{e,\schlange{e}}D_{f,\schlange{f}}\quad\mbox{(TA:DF)}\quad\\
c(e)=c(\schlange{e}), c(f)=c(\schlange{f}),s(e_2)=r(e_1),
r(f)=c(e),s(f_2)=r(f_1),c(\schlange{e})=r(\schlange{f})
\end{eqnarray}
$\falg_{u1}$ will be involutive if we have in addition
\begin{equation}
\sum_{e,f}D_{e,f}^\ast\schlange{\eta}_{e,e^\ast}\schlange{\eta}_{f,f^\ast}
=D_{e^\ast,f^\ast} \qquad\mbox{(TA:DN)}
\end{equation}
\end{de}

The commutator relations between this fields and the imbeddings
are straightforward:
$(\varphi\in\hilb,v^C\in V^C,h_C\in\hilb_C,\varphi\in\hilb)$
\begin{de}\begin{eqnarray}
\lefteqn{i_\eich(g)\Psi^C(h_C,v^C)(\varphi):=}\nonumber\\
&&\sum_{e,f,C=c(e)=c(f)}D_{e,f}\phi(e)
(h_C\otimes P^\oalg_{s(e)}(\varphi))\otimes
\varrho^{r(f)}(g)\chi(f)(v^C\otimes P^\eich_{s(f)}(\varphi))=\\
&& \sum_{e,f,C=c(e)=c(f)}D_{e,f}\phi(e)(h_C\otimes
P^\oalg_{s(e)}(\varphi))\otimes
\chi(f)(\varrho^C\tprod\varrho^{s(f)})(g)
(v^C\otimes P^\eich_{s(f)})(\varphi))
\nonumber\\
\lefteqn{i_\oalg(A)\Psi^C(h_C,v^C)(\varphi):=}\nonumber\\
&&\sum_{e,f,C=c(e)=c(f)}D_{e,f}\pi_R(A)\phi(e)(h_C\otimes
P^\oalg_{s(e)}(\varphi))\otimes
\chi(f)(v^C\otimes P^\eich_{s(f)}(\varphi))=\\
&& \sum_{e,f,C=c(e)=c(f)}D_{e,f}\phi(e)(\pi_C\tprod\pi_{s(e)})(A)
(h_C\otimes P^\oalg_{s(e)}(\varphi))\otimes
\chi(f)(v^C\otimes P^\eich_{s(f)}(\varphi))\nonumber
\end{eqnarray}\end{de}

\begin{bem}\begin{enumerate}
\item The localization properties of $\Psi^C$ stem from that of
the $\oalg$-vertices $\phi(e)$ while the gauge transformation are
usually not localized.
\item It is usual (and possible without any modifications) to exclude
$i_\eich$ from the field algebra (They may be unwanted because they
can't be localized.). We included it to have a
construction that is totally symmetric between the
gauge and the observable algebra.
One can take for example $\eich=\oalg$. In conformal QFT it is tempting to
interpret the antichiral algebra as the gauge
algebra of the chiral algebra and
vice versa.

\end{enumerate}
\end{bem}

The braid relations in $\falg_{u1}$ involve a $\rmat$ matrix
which has nonnumeric entries in the general case of not
coassociative $\eich$.

\begin{satz}[Braiding] Assume that the fields
$\Psi^{C_2}(h_{C_2}\otimes e^{C_2}_{c_2})$ and $\Psi^{C_1}
(h_{C_1}\otimes e^{C_1}_{c_1})$ are localized so that their
$phi$ vertices obey braid relations with $B^+$.

For coassociative $\eich$ the following braid relations hold:
\begin{equation}
\Psi^{C_2}(h_{C_2}\otimes e^{C_2}_{c_2})\Psi^{C_1}
(h_{C_1}\otimes e^{C_1}_{c_1})
=\sum_{c'_2,c'_1} \rmat^{C_2,C_1;c'_2,c'_1}_{c_2,c_1}
\Psi^{C_1}(h_{C_1}\otimes e^{C_1}_{c'_1})\Psi^{C_2}
(h_{C_2}\otimes e^{C_2}_{c'_2})
\end{equation}

In the general case this becomes:
\begin{eqnarray}
\lefteqn{\Psi^{C_2}(h_{C_2}\otimes e^{C_2}_{c_2})
\Psi^{C_1}(h_{C_1}\otimes e^{C_1}_{c_1})
=}\nonumber\\
&&\sum_{l,l',\schlange{c_1},\schlange{c_2}}
\phi^{-1;C_2,C_1,\schlange{c_2},\schlange{c_1}}_{l,c_2,c_1}
\sum_{c'_2,c'_1} \rmat^{c(e_2),c(e_1);c'_2,c'_1}_{\schlange{c_2},
\schlange{c_1}}
\sum_{\schlange{c'}_1,\schlange{c'}_2}\phi_{l',c'_1,c'_2}
                        ^{C_1,C_2,\schlange{c'_1},\schlange{c'_2}}\\
&&\Psi^{C_1}(h_{C_1}\otimes e^{C_1}_{\schlange{c'}_1})\Psi^{C_2}
(h_{C_2}\otimes e^{C_2}_{\schlange{c'}_2})
i_\eich(\phi_l^{(3)}\phi_{l'}^{(3)})=\\
&=&\sum_{c'_1,c'_2}
\Psi^{C_1}(h_{C_1}\otimes e^{C_1}{c'_1})
\Psi^{C_2}(h_{C_2}\otimes e^{C_2}_{c'_2})
(\varrho^{C_2,c'_2}_{c_2}\otimes\varrho^{C_1,c'_1}_{c_1}\otimes i_\eich)
(\phi^{-1;2,1,3}(R\otimes1)\phi)
\end{eqnarray}
With $\phi=\sum_l\phi_l^{(1)}\otimes\phi_l^{(2)}\otimes\phi_l^{(3)}$
and $\phi^{C_2,C_1,\schlange{c_2},\schlange{c_1}}_{l,c_2,c_1}
=\varrho^{C_2}(\phi^{(1)}_l)^{\schlange{c_2}}_{c_2}
\varrho^{C_1}(\phi_l^{(2)})^{\schlange{c_1}}_{c_1}$.
\end{satz}

\begin{bew}
\begin{eqnarray*}
\lefteqn{\Psi^{C_2}(h_{C_2}\otimes e^{C_2}_{c_2})
\Psi^{C_1}(h_{C_1}\otimes e^{C_1}_{c_1})
(h\otimes e^S_s)=}\\
&&\sum_{e_1,e_2,f_1,f_2}D_{e_1,f_1}D_{e_2,f_2}
\phi(e_2)(h_{C_2}\otimes\phi(e_1)(h_{C_1}\otimes h))\otimes\chi(f_2)
(e^{C_2}_{c_2}\otimes\chi(f_1)(e^{C_1}_{c_1}\otimes e^S_s))=\\
&=&\sum_{e'_1,e'_2,e_1,e_2,f_1,f_2}D_{e_1,f_1}D_{e_2,f_2}
B^+_{e_1,e_2;e'_2,e'_1}
\phi(e'_1)(h_{C_1}\otimes\phi(e'_2)(h_{C_2}\otimes h))\otimes\\
&&\sum_{q,r}\clebsch{C_1}{S}{Q}{c_1}{s}{q}
\clebsch{C_2}{Q}{R}{c_2}{q}{r}e^R_r=\\
&=&\sum_{e'_1,e'_2,r(e_1),r(e_2),r(f_1),r(f_2)}D_{e_1,f_1}D_{e_2,f_2}
B^+_{e_1,e_2;e'_2,e'_1}
\phi(e'_1)(h_{C_1}\otimes\phi(e'_2)(h_{C_2}\otimes h))\otimes\\
&&\sum_{P,p,c'_1,c'_2,\schlange{c_1},\schlange{c_2},\schlange{s}}
  \rmat^{C_2,C_1;c'_2,c'_1}_{\schlange{c_2},\schlange{c_1}}
\phi^{-1;C_2,C_1,S,\schlange{c_1},\schlange{c_2},\schlange{s}}_{c_2,c_1,s}
     \sum_{\schlange{c'}_1,\schlange{c'}_2,\schlange{s'}}
     \phi^{C_1,C_2,S,\schlange{c'}_1,\schlange{c'}_2,\schlange{s}'}_
       {c'_1,c'_2,\schlange{s}} \\
&&B^-_{f_1,f_2;f'_2,f'_1}
\clebsch{C_2}{S}{P}{\schlange{c'}_2}{\schlange{s'}}{p}
\clebsch{C_1}{P}{R}{\schlange{c'}_1}{p}{r}
e^R_{r}=\\
&=&\sum_{e'_1,e'_2,}D_{e'_1,f'_1}D_{e'_2,f'_2}
\phi(e'_1)(h_{C_1}\otimes\phi(e'_2)(h_{C_2}\otimes h))\otimes
\sum_{P,p,c'_1,c'_2,\schlange{c_1},\schlange{c_2},\schlange{s}}
  \rmat^{C_2,C_1;c'_2,c'_1}_{\schlange{c_2},\schlange{c_1}}	 \\
&&\phi^{-1;C_2,C_1,S,\schlange{c_1},\schlange{c_2},\schlange{s}}_{c_2,c_1,s}
     \sum_{\schlange{c'}_1,\schlange{c'}_2,\schlange{s'}}
     \phi^{C_1,C_2,S,\schlange{c'}_1,\schlange{c'}_2,\schlange{s}'}_
       {c'_1,c'_2,\schlange{s}}
\clebsch{C_2}{S}{P}{\schlange{c'}_2}{\schlange{s'}}{p}
\clebsch{C_1}{P}{R}{\schlange{c'}_1}{p}{r}
e^R_{r}=\\
&=&\sum_{l,l',\schlange{c_1},\schlange{c_2}}
\phi^{-1;C_2,C_1,\schlange{c_2},\schlange{c_1}}_{l,c_2,c_1}
\sum_{c'_2,c'_1} \rmat^{c(e_2),c(e_1);c'_2,c'_1}_{\schlange{c_2},
\schlange{c_1}}
\sum_{\schlange{c'}_1,\schlange{c'}_2}\phi_{l',c'_1,c'_2}
                        ^{C_1,C_2,\schlange{c'_1},\schlange{c'_2}}\\
&&\Psi^{C_1}(h_{C_1}\otimes e^{C_1}_{\schlange{c'}_1})\Psi^{C_2}
(h_{C_2}\otimes e^{C_2}_{\schlange{c'}_2})
i_\eich(\phi_l^{-1;(3)}\phi_{l'}^{(3)})(h\otimes e^S_s)
\end{eqnarray*}
With $Q:=r(f_1),R:=r(f_2)$. The third step used the vertex sos
transformation, the fourth used equation (\ref{ddef}).
\end{bew}

\begin{bem}
The operators in $\falg_{u1}$
form a representation of the quantum plane.
\end{bem}

\begin{satz}[Fusion]	\label{fus1}
For ($h\in\hilb_S$) the fusion reads
\begin{eqnarray}
\lefteqn{\Psi^{C_2}(h_{C_2}\otimes e^{C_2}_{c_2})
         \Psi^{C_1}(h_{C_1}\otimes e^{C_1}_{c_1})
(h\otimes e^S_s)=
\sum_{e,f} \sum_{c'_2,c'_1,s'}\phi^{-1;C_2,C_1,S,c'_2,c'_1,s'}_{c_2,c_1,s}}\\
&=&
\Psi^{c(e)}\left((P^\oalg_{c(e)}\otimes P^\eich_{c(e)})\Psi^{c(f)}
(h_{C_2}\otimes e^{C_2}_{c'_2})(h_{C_1}\otimes e^{C_1}_{c'_1})\right)
(h\otimes e^S_{s'})\nonumber
\end{eqnarray}
\end{satz}
\begin{bew}
\begin{eqnarray*}
\lefteqn{\Psi^{C_2}(h_{C_2}\otimes e^{C_2}_{c_2})
         \Psi^{C_1}(h_{C_1}\otimes e^{C_1}_{c_1})
(h\otimes e^S_s)=}\\
&=&\sum_{e_1,e_2,f_1,f_2}D_{e_1,f_1}D_{e_2,f_2}
\phi(e_2)(h_{C_2}\otimes\phi(e_1)(h_{C_1}\otimes h))\otimes
\chi(f_2)(e^{C_2}_{c_2}\otimes\chi(f_1)(e^{C_1}_{c_1}\otimes e^S_s))=\\
&=&\sum_{e_1,e_2,f_1,f_2}D_{e_1,f_1}D_{e_2,f_2}
\sum_{e,f,\schlange{e},\schlange{f}}
F_{e_1,e_2;f,e}F_{f_1,f_2;\schlange{f},\schlange{e}}
\phi(e)(\phi(f)(h_{C_2}\otimes h_{C_1})\otimes h)\otimes\\
&&\sum_{c'_2,c'_1,s'}\phi^{-1;C_2,C_1,S,c'_2,c'_1,s'}_{c_2,c_1,s}
\chi(\schlange{e})(\chi(\schlange{f})(e^{C_2}_{c'_2}\otimes e^{C_1}_{c'_1})
\otimes e^S_{s'})=\\
&=&\sum_{e,f} \sum_{c'_2,c'_1,s'}\phi^{-1;C_2,C_1,S,c'_2,c'_1,s'}_{c_2,c_1,s}
\Psi^{c(e)}((P^\oalg_{c(e)}\otimes P^\eich_{c(e)})\Psi^{c(f)}
(h_{C_2}\otimes e^{C_2}_{c'_2})(h_{C_1}\otimes e^{C_1}_{c'_1}))
(h\otimes e^S_{s'})
\end{eqnarray*}
\end{bew}

\begin{satz} $\falg_{u_1}$ is closed under taking adjoints.
\end{satz}
\begin{bew}
Using  TA:DN and (\ref{konjform}) we find:
\begin{eqnarray*}
\Psi^C(h_C\otimes v^C)^\ast
&=&\sum_{e,f}\sum_{e^\ast,f^\ast}D^\ast_{e,f}\schlange{\eta}_{e,e^\ast}
\schlange{\eta}_{f,f^\ast}
\phi(e^\ast)(\schlange{h_C}\otimes\cdot)\otimes
\chi(f^\ast)(\schlange{v^C}\otimes\cdot)\\
&=&\sum_{e^\ast,f^\ast}D_{e^\ast,f^\ast}
\phi(e^\ast)(\schlange{h_C}\otimes\cdot)\otimes
\chi(f^\ast)(\schlange{v^C}\otimes\cdot)\\
&=&\Psi^{C^\ast}(\schlange{h_C}\otimes\schlange{v^C})
\end{eqnarray*}
\end{bew}

\begin{bem} \label{konjbem009}
The proof shows that adjoint field operators transform
according to the conjugate representation.
\end{bem}

\sbegin{bem}[Covariant Operator Products]
The fusion and braiding relations proved sofar
are ugly since they involve nonnumerical matrices.
Mack/Schomerus showed how to repair this unsatisfactory situation by
absorbing this operators in the definition
of a covariant operator product.
\begin{equation}
\Psi^{C_2}(h_{C_2}\otimes v^{C_2})\times\Psi^{C_1}(h_{C_1}\otimes v^{C_1})
:= \sum_l
\Psi^{C_2}(h_{C_2}\otimes\varrho^{C_2}(\phi^{(1)}_l)v^{C_2}))
\Psi^{C_1}(h_{C_1}\otimes\varrho^{C_1}(\phi^{(2)}_l)v^{C_1}))
i_\eich(\phi^{(3)}_l)
\end{equation}
This product is not associative. Altering the parentheses
yields conjugation by $i_\eich(\phi)$.

Fusion and braiding now look like
\begin{eqnarray}
\lefteqn{\Psi^{C_2}(h_{C_2}\otimes e^{C_2}_{c_2})
   \times\Psi^{C_1}(h_{C_1}\otimes e^{C_1}_{c_1})
(h\otimes e^S_s)=}\nonumber\\
&=&\sum_{e,f}
\Psi^{c(e)}((P^\oalg_{c(e)}\otimes P^\eich_{s(e)})\Psi^{c(f)}
(h_{C_2}\otimes e^{C_2}_{c_2})(h_{C_1}\otimes e^{C_1}_{c_1}))
(h\otimes e^S_{s})\\
\lefteqn{\Psi^{C_2}(h_{C_2}\otimes e^{C_2}_{c_2})\times
\Psi^{C_1}(h_{C_1}\otimes e^{C_1}_{c_1})
=}\nonumber\\
&&
\sum_{c_2,c_1} \rmat^{c(e_2),c(e_1);c_2,c_1}_{\schlange{c_2},\schlange{c_1}}
\Psi^{C_1}(h_{C_1}\otimes e^{C_1}_{\schlange{c}_1})\times\Psi^{C_2}
(h_{C_2}\otimes e^{C_2}_{\schlange{c}_2})
\end{eqnarray}
\send{bem}

\begin{bem} We used freely $\phi^{-1}$ although $\phi$ needs not be invertible.
But it always has a quasi inverse $\phi\phi^{-1}=(id\otimes\Delta)\Delta(1),
\phi^{-1}\phi=(\Delta\otimes id)\Delta(1)$.

The resulting factors are harmless:
$\Psi^I(h\otimes v)=i_\eich(1)\Psi^I(h\otimes v)=
\Psi^I(h\otimes\varrho^I(1^{(1)}_l)v)i_\eich(1^{(2)}_l)$ with
$\Delta(1)=\sum_l 1^{(1)}_l\otimes 1^{(2)}_l$.
This shows: $\Psi^I(h_I\otimes v^I)\Psi^J(h_J\otimes v^J)=
i_\eich(1)\Psi^I(h_I\otimes v^I)\Psi^J(h_J\otimes v^J)=
\Psi^I(h_I\otimes\varrho^I(\cdot)v_I)
\Psi^I(h_J\otimes\varrho^J(\cdot)v_J)
i_\eich(\cdot)((id\otimes\Delta)\Delta(1))$.
\end{bem}

\begin{bem}
The $\Psi$ may be further specialised by setting
\begin{equation}             \label{gamma67}
\Gamma^I_i:=\Psi^I(h_I\otimes e^I_i)
\end{equation}
Where $h_I$ is the highest weight vector in $\hilb_I$.

Writing $\Delta(g)=\sum_l g^{(1)}_l\otimes g^{(2)}_l$
the transformation rule becomes
\begin{equation}
i_\eich(g)\Gamma^I_i=\sum_{l,k}\Gamma^I_k
  \varrho^I(g^{(1)}_l)_{i,k}i_\eich(g_l^{(2)})
\end{equation}
This is the form postulated by Mack/Schomerus.
\end{bem}

\begin{bem}
Correlations ${\langle} 0 {\mid}
\Psi^{C_n}(h_{C_n}\otimes v^{C_n})\circ\ldots\circ
\Psi^{C_1}(h_{C_1}\otimes v^{C_1})\vak,
h_{C_i}\in\hilb_{C_i},v^{C_i}\in V^{C_i}$ transform
covariantly under the gauge algebra.
If the trivial representation occurs in the reduction of
$\varrho^{C_n}\tprod\ldots\tprod\varrho^{C_1}$
this correlation may be gauge invariant. This is the case iff
$v^{C_n}\otimes\ldots\otimes v^{C_1}$ is mapped to a trivial representaion
via the reduction isomorphism.
Such invariant correlations are called conformal blocks in CQFT. In this
language our result is the same as \cite{gomez2}[(5.19)].
\end{bem}

\subsection{Field Algebra $\falg_{u2}$}

The construction of $\falg_{u1}$ depends on two technical
axioms. It is possible to alter the construction of
$\falg_{u1}$ in such a way that these axioms are at least in the case of AQFT
always satisfied.

The starting point is the following observation:
Rigid braided tensor categories are involutive.
To every isomorphism ${\rm Rep}(\oalg)\iso
{\rm Rep}(\eich)$ there is a second one
defined by an additional involution (Since we are
mainly interested in AQFT we write
${\quer{I}}$ instead of $I^\ast$.).

$\falg_{u2}$ operates on
$\hilb:=\bigoplus_I\hilb^I\otimes V^{\quer{I}}$.
$P^\oalg_I:\hilb\pfeil\hilb_I,
P^\eich_I:\hilb\pfeil V^{\quer{I}}$  denote the natural projections.
$\falg_{u2}$ is generated by
$\eich$, $\oalg$ and intertwiners.
We have natural embeddings $i_\oalg:\oalg\pfeil\falg_{u2}$ and
$i_\eich:\eich\pfeil\falg_{u2}$ operating on $h_I\otimes v^{\quer{I}}\in
\hilb_I\otimes V^{\quer{I}}$
as $i_\oalg(A)(h_I\otimes v^I):=\pi_I(A)h_I\otimes v^I$ and
$i_\eich(g)(h_I\otimes v^{\quer{I}}):=
h_I\otimes \varrho^{\quer{I}}(g)v^{\quer{I}}$.
$i_\eich$ and $i_\oalg$ commute.

\begin{de} $\falg_{u2}$ is generated by
 $i_\oalg(\oalg),i_\eich(\eich)$ and intertwiners:
\begin{eqnarray}
\lefteqn{\Psi^C(h_C\otimes v^{\quer{C}}):\hilb\pfeil\hilb}
\end{eqnarray}\begin{eqnarray}
\Psi^C(h_C\otimes v^{\quer{C}}):=
\sum_{e,\quer{e},c(e)=C,c(\quer{e})={\quer{C}}}
D_{e,\quer{e}}
\phi(e)(h_C\otimes \cdot)\otimes\chi(\quer{e})(v^{\quer{C}}\otimes \cdot)
\label{psidef22}\\
v^{\quer{C}}\in V^{\quer{C}}, h_C\in\hilb_C
\end{eqnarray}

$D:W\otimes W\pfeil\CC,(e,f)\mapsto D_{e,f}\in\CC$ must satisfy:
\begin{eqnarray}
\sum_{r(e_1),r(\quer{e}_1)} B^+_{e_1,e_2;e'_2,e'_1}
B^-_{\quer{e}_1,\quer{e}_2;\quer{e}'_2,\quer{e}'_1}
D_{e_1,\quer{e}_1}D_{e_2,\quer{e}_2}
=D_{e'_1,\quer{e}'_1}D_{e'_2,\quer{e}'_2}\quad\mbox{(TA:DB)}\quad\\
c(e_1)=c(\quer{e}_1),c(e_2)=c(\quer{e}_2),s(e_2)=r(e_1),
s(e'_1)=r(e'_2),s(\quer{e}_2)=r(\quer{e}_1),s(\quer{e}'_1)=r(\quer{e}'_2)
\end{eqnarray}
\begin{eqnarray}
\sum_{r(e_1),r(\quer{e_1})} F_{e_1,e_2;f,e}
F_{\quer{e_1},\quer{e_2};\quer{f},\quer{e}}
D_{e_1,\quer{e_1}}D_{e_2,\quer{e_2}}
=D_{e,\quer{e}}D_{f,\quer{f}}\quad\mbox{(TA:DF)}\quad\\
c(e)=c(\quer{e}), c(f)=c(\quer{f}),s(e_2)=r(e_1),
r(f)=c(e),s(\quer{e_2})=r(\quer{e_1}),c(\quer{e})=r(\quer{f})
\end{eqnarray}
$\falg_{u2}$ will be involutive if:
\begin{equation}
\sum_{e,\quer{e}}D_{e,\quer{e}}^\ast\schlange{\eta}_{e,e^\ast}
\schlange{\eta}_{\quer{e},\quer{e}^\ast}
=D_{e^\ast,\quer{e}^\ast}\qquad\mbox{(TA:DN)}
\end{equation}
\end{de}

In algebraic  QFT there are alway solutions:
$D_{e,\quer{e}}:=\zeta_{e,\quer{e}}$.
This setting transforms TA:DF and TA:DB to well known identities
in AQFT (see \cite{rehren4}). TA:DN can also be reduced to a
standard formula (\cite{rehren7}) by bringing the second
$\eta$ to the right by means of orthogonality.

\section{Was it all worth it?}

Is the generality of weak quasi quantum groups really needed or can one do
with ordinary quantum groups? We split our answer in two parts:

\subsection{Truncation is unavoidable}\label{hks}

The (chiral) observable algebra of minimal conformal models is
just the Virasoro algebra.
The simplest example is the Ising model with fusion rules
$\sigma\times\sigma=1+\epsilon,\sigma\times\epsilon=
\epsilon\times\sigma=\sigma,
\epsilon\times\epsilon=1$.
Assume the dimension of the representations of the symmetry
algebra to be $D(1)=1,D(\sigma)=n,D(\epsilon)=m\in\NN$. We deduce
$n^2=1+m,nm=n\Rightarrow m=1,n=\sqrt{2}$. This shows that there can't be a
dimension function for the Ising model. Truncation have to take place leaving
only room for weak dimension functions.

This kind of argument was generalized by H. Kratz to all minimal models with
multiplicities $N_{I,J}^K\in\{0,1\}$ \cite{kratz}.

\subsection{Weak Quasi Hopf algebras are unavoidable}\label{nogo}

Mack and Schomerus \cite{ms0} have shown that an ordinary quantum group
as symmetry algebra contradicts with the existence of braid relations on
all of $\hilb$.

Here is an argument that works also in the general case treated in this paper:
Consider the fusion (proposition \ref{fus1}) in the case of an
ordinary quantum group (i.e. $\phi$ trivial) and assume truncation of some
unphysical representations to be carried out by hand.
Apply the fusion formula to $\Psi^{C_0}(h_{C_0}\otimes v^{C_0})\vak$, where
$v^{C_0}$ and the quantum group vector $v_1\in V^{C_1}$ of the second
operator are chosen so that their tensor product is unphysical.
Then the lefthand side of the equation is zero because of truncation.
However the righthand side will not allways vanish: We can set
$C_2:=C_1^\ast$ and
by rigidity we can find a
$v_2\in V^{C_2}$ such that its tensor product with $v_1$
will not vanish.
Therefore the right hand side gets a contribution in the $C_0$ sector.

\section{Remarks and open questions}

\begin{enumerate}
\item In contrast to the classical case analysed by Doplicher and Roberts
neither the weak dimension function, the weak quasi tensor functor,
the weak quasi hopf algebra
nor the covariant field algebra are determined uniquely.
This is strange because the dimension of the gauge representations
can in principle be determined by measurement:
They give the size of field multiplets
and thereby determine the amplitude of particle creation processes.
\item Describe ultra weak quasi Hopf algebras broken gauge symmetries?
\item Every solution of Moore/Seiberg's equations
yields a semisimple braided tensor
category. In analogy to proposition \ref{funktorkon} one can build an
ultraweak quasi tensor functor from the assignment of a fixed separable
hilbert space to every irreducible object:  $F(X):=\hilb_0$.
We expect the reconstructed ultraweak quasi Hopf algebra $H$
to play the role of the Virasoro or Kac-Moody algebras
in chiral conformal QFT.
Guided by the idea "QFT = category+manifold" we ask if for every manifold M
there exists  a  sort of parallel transport
$\delta_{x,y}:H\pfeil H$, $x,y\in M$
compatible with some space time symmetry such that one can glue
the copies of $H$ associated to every point in $M$ to a full (chiral)
observable algebra.\\[0.4cm]
\end{enumerate}
{\em\bf Acknowledgement:} It is a pleasure to thank all those who
helped in the trail of this work.
I thank my fellow students Henrik Kratz
(who contributed to parts of this paper
by careful analysis and criticism), Andreas Schmidt and Wolfram Boenkost
for discussions and
Professor F. Constantinescu for his continous interest in this work.

G. Mack gave introduction to weak quasi hopf algebras, K.-H. Rehren
explained details of his work, S. Majid and J. E. Roberts
gave valuable criticism.

Thanks to the Studienstiftung des deutschen Volkes for
financial support and more.

Almost nothing of the theory presented here is really new. In fact this
paper is nothing more than a footnote to the work of
Kerler, Mack/Schomerus and Majid.

\twocolumn

\end{document}